\documentclass[structabstract]{aa}  
\usepackage{graphicx}
\usepackage{amsmath}
\usepackage{txfonts}
\usepackage{natbib}
\usepackage{nameref}                    % References
\usepackage{color}
\usepackage[colorlinks,citecolor=blue,linkcolor=red,urlcolor=cyan]{hyperref}
\newcommand{\teff}  {T$_\mathrm{eff}$}
\newcommand{\logg}  {$\log g$}

\def\aap{A\&A}

\def\apjs{ApJS}
\def\apjl{ApJL}

\begin{document}

   \title{Chemical abundances of 1111 FGK stars from the HARPS GTO
planet search program IV.}

\subtitle{Carbon and C/O ratios for Galactic stellar populations and planet hosts\thanks{
Based on observations collected at the La Silla Observatory, ESO
(Chile), with the HARPS spectrograph at the 3.6 m ESO telescope (ESO
runs ID 72.C--0488, 082.C--0212, and 085.C--0063).}\thanks{The table with parameters and abundances is only available in electronic form
at the CDS via anonymous ftp to cdsarc.u-strasbg.fr (130.79.128.5)
or via http://cdsweb.u-strasbg.fr/cgi-bin/qcat?J/A+A/}}

   \author{E. Delgado Mena\inst{1}
      \and V. Adibekyan\inst{1}
      \and N.~C.~Santos\inst{1,2}
      \and M. Tsantaki\inst{3}
      \and J.~I.~Gonz\'alez Hern\'andez\inst{4,5}
      \and S. G. Sousa\inst{1,2}
      \and S. Bertr\'an de Lis\inst{6}
      }
      
\institute{
Instituto de Astrof\'isica e Ci\^encias do Espa\c{c}o, Universidade do Porto, CAUP, Rua das
Estrelas, PT4150-762 Porto, Portugal
             \email{Elisa.Delgado@astro.up.pt}
\and
Departamento de F\'isica e Astronom\'ia, Faculdade de Ci\^encias, Universidade do Porto, Portugal
\and 
INAF -- Osservatorio Astrofisico di Arcetri, Largo E. fermi 5, 50125 Firenze, Italy
\and
Instituto de Astrof\'{\i}sica de Canarias, C/ Via Lactea, s/n, 38205, La Laguna, Tenerife, Spain %\\
\and 
Departamento de Astrof\'isica, Universidad de La Laguna, 38206 La Laguna, Tenerife, Spain
\and
The Johns Hopkins University, Baltimore, USA
}

%             University of Alexandria, Department of Geography, ...\\
%             \email{c.ptolemy@hipparch.uheaven.space}
%             \thanks{The university of heaven temporarily does not accept e-mails}

   \date{Received ...; accepted ...}

% \abstract{}{}{}{}{} 
% 5 {} token are mandatory
 
  \abstract
{}
{We study the carbon abundances with a twofold objective. On the one hand, we want to evaluate the behaviour of carbon in the context of Galactic chemical evolution. On the other hand, we focus on the possible dependence of carbon abundances on the presence of planets and on the impact of various factors (such as different oxygen lines) on the determination of C/O elemental ratios.}
{We derived chemical abundances of carbon from two atomic lines for 757 FGK stars in the HARPS-GTO sample, observed with high-resolution ($R \sim$\,115000) and high-quality spectra. The abundances were derived with the code MOOG using automatically measured EWs and a grid of Kurucz ATLAS9 atmospheres. Oxygen abundances, derived using different lines, were taken from previous papers in this series and updated with the new stellar parameters.}
{We find that thick- and thin-disk stars are chemically disjunct for [C/Fe] across the full metallicity range that they have in common. Moreover, the population of high-$\alpha$ metal-rich stars also presents higher and clearly separated [C/Fe] ratios than thin-disk stars up to [Fe/H]\,$\sim$\,0.2\,dex. The [C/O] ratios present a general flat trend as a function of [O/H] but this trend becomes negative when considering stars of similar metallicity. We find tentative evidence that stars with low-mass planets at lower metallicities have higher [C/Fe] ratios than stars without planets at the same metallicity, in the same way as has previously been found for $\alpha$ elements. Finally, the elemental C/O ratios for the vast majority of our stars are below 0.8 when using the oxygen line at 6158$\AA,{}$ however, the forbidden oxygen line at 6300$\AA{}$ provides systematically higher C/O values. Moreover, by using different atmosphere models the C/O ratios can have a non negligible difference for cool stars. Therefore, C/O ratios should be scaled to a common solar reference in order to correctly evaluate its behaviour. We find no significant differences in the distribution of C/O ratios for the different populations of planet hosts, except when comparing the stars without detected planets with those hosting Jupiter-type planets. However, we note that this difference might be caused by the different metallicity distributions of both populations.
}
{The derivation of homogeneous abundances from high-resolution spectra is of great utility in constraining models of GCE. This high-quality data combined with the long-term study of planetary presence in our sample is crucial for achieving an accurate understanding of the impact of stellar chemical composition on planetary formation mechanisms.}


\maketitle
%
%________________________________________________________________

\section{Introduction}

The determination of stellar carbon abundances is of prime interest for a range of topics in astrophysics, such as the chemical evolution of the Galaxy \citep{carigi05,kobayashi20}, the properties of carbon-rich dwarfs \citep{farihi18}, and the composition and structure of exoplanets \citep{bond10,dorn15}. Despite being one of the most abundant elements in the Universe, the formation of carbon is not yet fully understood. This problem is made particularly difficult for a number of reasons. For instance, from the observational point of view, different works in the literature find distinct trends for its evolution across metallicity \citep[e.g.][]{bensby06,franchini20}. This might be caused by the determination of carbon abundances from different carbon indicators (either atomic or molecular lines), which may provide contradictory results in some cases \citep[e.g.][]{bensby06,suarez-andres17,franchini20}. Moreover, the consideration of Non-Local Thermodinamyc Equilibrium (NLTE) and 3D effects can have a substantial impact on the evolution of carbon at very low metallicities \citep{amarsi19}. On the other hand, from the theoretical point of view, the carbon yields from massive as well as from low and intermediate-mass stars depend on a number of factors (e.g. stellar winds, convection treatment), which need to be accounted for in the Galactic chemical evolution (GCE) models in order to aptly reproduce the observations \citep{carigi05}.

By studying different populations of stars at different metallicities, we can understand which processes have played a major role in the production of different elements at a given moment of the evolution of the Galaxy, providing constrains for the current models of GCE. Furthermore, carbon is an essential element for life (in the forms that are familiar to us on Earth) and, as such, there is a clear interest in the study of whether stars hosting planets present different carbon content. It is also important for us to understand  whether this element has an impact on the formation of planetary systems.

The objective of this work is to derive carbon abundances for the HARPS-GTO sample using atomic carbon lines to study the GCE of this element and the possible relation between the abundance ratios [C/Fe] and C/O and the presence of planets. We note that carbon abundances obtained from the CH band were derived for this sample in \citet{suarez-andres17} (however, based on an older set of stellar parameters) and, hence, we can also compare the results from different carbon indicators. Moreover, our volume-limited sample contains a significant number of metal rich stars ([Fe/H]\,$>$\,0.2\,\,dex) which permits us to study the GCE at high metallicities, which is not often explored in the literature. 

Over the last decade, our team has dedicated extensive and detailed studies to the chemical characterisation of the HARPS-GTO sample. Chemical abundances for refractory elements with atomic number $Z$\,$<$\,29 can be found in \citet{adibekyan12}, whereas the heavier elements ($Z$\,$\geq$\,29) are published in \citet[][hereafter DM17]{delgado17} and sulfur in \citet{costasilva20}. We also derived abundances for the following light elements in the HARPS-GTO sample: oxygen \citep{bertrandelis15}, carbon from the CH band \citep{suarez-andres17}, and lithium \citep{delgado14,delgado15}. Finally, for a small fraction of the stars, we made use of near-UV spectra obtained with UVES to derive the beryllium \citep{santos_Be02,santos_be_plan,galvez11,delgado_Be1,delgado_Be2} and nitrogen \citep{suarez-andres16} abundances.

This paper is organised as follows. In Sect. \ref{sec:obs}, we briefly describe the collected data and stellar parameters. In Sect. \ref{sec:abundances}, we detail the derivation of abundances and the comparison to other carbon indicators. In Sect. \ref{sec:CFe_pop}, we discuss the behaviour of different carbon abundance ratios in the context of the chemical evolution in the Galaxy and in Sect. \ref{sec:CFe_planets}, we analyse the behaviour of carbon and oxygen abundance ratios in the groups of stars with and without detected planets. Moreover, in Sect. \ref{sec:CO_variation}, we analyse the impact of different stellar atmosphere models and oxygen indicators on the determination of C/O ratios. Finally, we present our conclusions in Sect. \ref{sec:summary}.

\section{Observations and stellar parameters} \label{sec:obs}

The baseline sample used in this work consists of 1111 FGK stars observed within the context of the HARPS GTO planet search programs \citep{mayor03,locurto,santos_harps4}. It is a volume-limited sample (with most of the stars closer than 60 pc), where objects that are not optimal for planet searches (such as fast rotators or active stars) were removed. The HARPS spectra have a resolution of R $\sim$115000 and covers the range 3770-6900$\AA{}$. It is reduced with the specific instrument pipeline. Our sample stars have been observed multiple times between 2003 and 2010 in the context of planet search programs. The individual spectra of each star is combined with IRAF to obtain a higher S/N final spectra (45$\%$ of the spectra have 100\,$<$\,S/N\,$<$\,300, 40$\%$ of the spectra have S/N\,$>$\,300, and the mean S/N is 380 as measured in the region 5700-6000$\AA{}$).

Precise stellar parameters for the full sample of 1111 stars within the HARPS-GTO program were first homogeneously derived in \cite{sousa08,sousa_harps4,sousa_harps2} by using the excitation and ionization balance method on a set of \ion{Fe}{I} and \ion{Fe}{II} lines for which the equivalent widths (EWs) are measured with ARES \citep{sousa_ares} and the abundances are determined with the radiative transfer code MOOG \citep{sneden}. The parameters for cool stars of a fraction of the sample were revised by \citet{tsantaki13} using the same method but with a list of iron lines that are more adequate for cool stars. This method for cool stars was later applied to the full HARPS-GTO sample in DM17. Moreover, the spectroscopic gravities were corrected by using trigonometric gravities derived from parallaxes for all the stars. We refer the reader to DM17 for a figure showing the Kiel diagram of the full sample. Our stars have typical \teff\ values between 4500\,K and 6500\,K and surface gravities mostly lie in the range 4\,$<$\,$\log g$\,$<$\,5 \,dex, meanwhile the metallicity covers the region -1.39\,$<$\,[Fe/H]\,$<$\,0.55\,\,dex. In a subsequent work, \citet{delgado19}, we derived stellar ages for the full sample, by using \textit{Gaia} DR2 parallaxes and PARSEC isochrones \citep{bressan12}.

The total sample is composed by 152 stars with planets (29 of them hosting only Neptunians or Super-Earths: M\,$<$\,30\,M$_{\oplus}$) and 959 stars without detected planets (hereafter, single stars\footnote{We note that due to the precision of radial velocities achieved with HARPS and the long-term survey, we can be sure that giant planets do not exist around such stars but we cannot rule out the presence of small planets, especially at long periods.}).

\begin{center}
\begin{table}
% use packages: array
\caption{Atomic parameters for the lines used in this work together with EWs and absolute abundances in the Sun from the Kurucz ATLAS solar spectrum and considering the \logg\ corrected values\tablefootmark{*}.}
\label{lineas}
\centering
%\resizebox{\linewidth}{!}{%
\begin{tabular}{lccrcc}
\hline
\noalign{\medskip} 
element & $\lambda$ ($\AA{}$) & $\chi_{l}$ (eV) & log \textit{gf} & EW(m$\AA{}$)  & A(X)$_{\odot}$ \\
\noalign{\medskip} 
\hline
\hline
\noalign{\smallskip} 
\ion{C}{I} & 5052.15 &  7.68 &   --5.166 &  34.5 & 8.466 \\
\ion{C}{I} & 5380.32 &  7.68 &   --1.616 &  21.0 & 8.473 \\
\noalign{\smallskip} 
\hline
\noalign{\smallskip} 
\ion{O}{I} & 6158.17 & 10.74 &   --0.296 &  3.7 & 8.722 \\
$[\ion{O}{I}]$ & 6300.30 &  0.00 &   --9.717 &  3.4 & 8.635 \\
\ion{Ni}{I} & 6300.33 & 4.27 &   --0.211 &  2.0 & 6.250  \\
\noalign{\smallskip} 
\hline
\noalign{\smallskip} 
\end{tabular}
%}
\tablefoottext{*}{Stellar parameters for the Sun are \teff\,=\,5777\,K, \logg\,=\,4.40 and [Fe/H]\,=\,0.0\,\,dex (A(Fe)=\,7.47\,dex). The last column corresponds to the absolute abundances.}
\end{table}
\end{center}

\begin{table*}
\centering
\caption{Average abundance sensitivities of carbon to changes of each parameter by their individual $\sigma$.}
\label{table_errors}
\begin{tabular}{lcccccccccc}
\hline
%\noalign{\vskip0.02\columnwidth}
 & \emph{continuum error} 
 & \emph{$\Delta$T$_\mathrm{eff}$} $=\pm$ $\sigma_{T_\mathrm{eff}}$ 
 & \emph{$\Delta$}{[}Fe/H{]}\emph{ }$=\pm$ $\sigma_{[Fe/H]}$ 
 & \emph{$\Delta$}$\log{g}$\emph{ }$=\pm$ $\sigma_{\log{g}}$ 
 & \emph{$\Delta$}$\xi_{\mathrm{t}}$ $=\pm$ $\sigma_{\xi_{\mathrm{t}}}$ \\
\hline 
\hline
\noalign{\smallskip} 
\noalign{\smallskip} 
low \emph{T$_\mathrm{eff}$}  & $\pm$0.038 & $\pm$0.027 & $\pm$0.001 & $\mp$0.022 & $\pm$0.000 \\
low \emph{solar}             & $\pm$0.016 & $\pm$0.013 & $\pm$0.001 & $\mp$0.010 & $\pm$0.000 \\
high \emph{T$_\mathrm{eff}$} & $\pm$0.017 & $\pm$0.017 & $\pm$0.002 & $\mp$0.015 & $\pm$0.002 \\
\noalign{\smallskip} 
\hline 
\end{tabular}
\end{table*}

\section{Derivation of carbon abundances} \label{sec:abundances}
Carbon abundances were derived from two well-known \ion{C}{I} lines (5052\AA{} and 5380\AA{}) under a standard LTE analysis with the 2017 version of the code MOOG \citep{sneden} using the \textit{abfind} driver. A grid of Kurucz ATLAS9 atmospheres \citep{kurucz} was used as the input, along with the EWs and the atomic parameters, wavelength ($\lambda$), excitation energy of the lower energy level ($\chi$), and oscillator strength (log \textit{gf}) of each line. The atomic parameters of the lines and the Van der Waals damping constants, log ($\gamma _{6}/N_{H}$), were retrieved from VALD3 database\footnote{http://vald.astro.univie.ac.at/$\sim$vald3/php/vald.php} \citep{vald15}. The EWs of the different lines were measured automatically with version 2 of the ARES program\footnote{The ARES code can be downloaded at http://www.astro.up.pt/~sousasag/ares/} \citep{sousa_ares2,sousa_ares}; however, for some stars, a visual inspection was needed to correctly measure the EWs. The atomic data, EWs and derived abundances for the Sun are shown in Table \ref{lineas}, where abundances are given in the classical form A(X) = log[(N(X)/N(H)] + 12.   

\begin{figure}
\centering
\includegraphics[width=9.0cm]{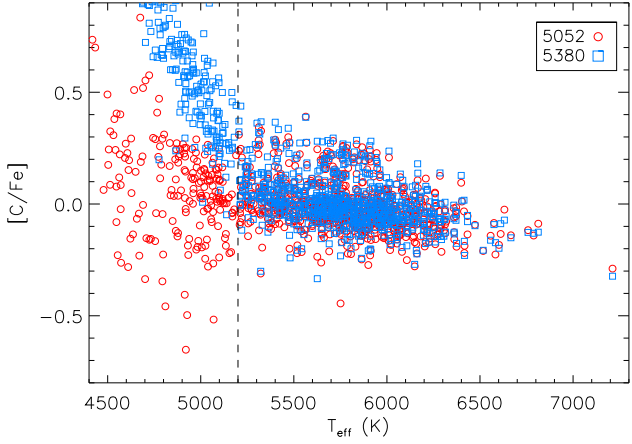}
\caption{[C/Fe] ratios for each line as a function of \teff.} 
\label{CFe_teff}
\end{figure}

\begin{figure}
\centering
\includegraphics[width=9.0cm]{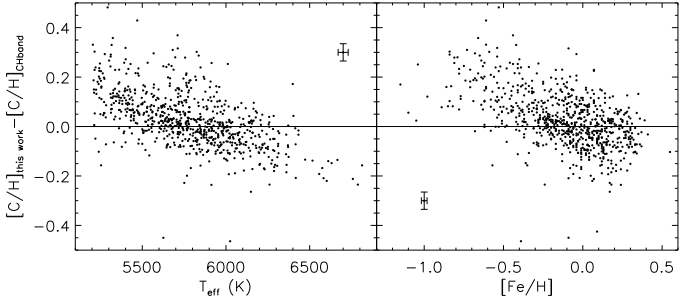}
\caption{Difference in [C/H] ratios between this work and the abundances from \citet{suarez-andres17} based on the CH band as a function of \teff and [Fe/H].} 
\label{comp_CH_band}
\end{figure}

Since we are also interested in the study of the C/O elemental ratio, we made use of the oxygen measurements previously derived for this sample by \cite{bertrandelis15}. However, those abundances were derived with the older set of parameters provided in \cite{sousa08,sousa_harps4,sousa_harps2}. Therefore, we rederived the abundances with the new set of parameters from DM17 and the published set of EWs in \cite{bertrandelis15}, for which the Ni blending line was removed. Since the abundances of oxygen can only be measured reliably for stars hotter than 5200\,K, the only difference in stellar parameters between this study and the original one is in the \logg\ because the other parameters only change for cooler stars. We note that \cite{bertrandelis15} considered the Kurucz Solar Atlas \citep{kurucz84} as a reference for solar oxygen abundances and thus we also use that solar spectrum to derive our carbon solar reference abundances. Since we corrected the spectroscopic \logg\ values for our full sample, we also applied that correction to our solar parameters (see Eq. 2 in DM17). 

The atomic lines we employ to derive abundances have a high excitation potential and, thus, they become weaker as \teff\ decreases, making it very difficult to derive reliable abundances for cool stars. Indeed, in Fig. \ref{CFe_teff}, a slight trend of increasing carbon abundances towards cooler temperatures can be observed, which becomes steeper at a \teff $\lesssim$ 5200\,K, especially for the line at 5380\AA{}, pointing to the possible presence of unknown blends. This line has also been reported to provide overabundances for cool and young stars that are also active \citep{baratella20}, although we note that the cool stars in our sample are all older than 4 Ga and are not active (since our sample was cleaned of active stars in order to ease the search for planets). Moreover, the abundances provided by both lines start to disagree around this value for \teff. As a comparison, the average standard deviation of both abundances for stars below 5200\,K is 0.47\,dex, whereas for the stars hotter than 5200\,K, it is below 0.05 (see the column marked 'continuum error' in Table \ref {table_errors}). Therefore, we set the limit for reliable abundances at 5200\,K, which is the same \teff\ cut applied to derive oxygen abundances by \cite{bertrandelis15} in the sample. In the end, we could derive reliable C abundances for 757 stars out of the 1111 that form the full sample. In Fig. \ref{CFe_teff}, we can also see that both carbon optical lines provide very similar abundances for stars above 5200\,K.

The abundance errors due to the uncertainty on the continuum placement are considered as the line-to-line scatter. The abundance errors due to the errors on stellar parameters were estimated by calculating the abundance differences when one of each of the stellar parameters was modified by its individual error. The average abundance sensitivities are shown in Table \ref{table_errors} for the same three groups of stars depending on \teff\ as done in \citet{adibekyan12} and DM17: ``low \teff'' stars -- stars with 5200\,K$<$ \teff\,$<$ 5277\,K, ``solar'' -- stars with \teff\,=\,T$_{\odot}$\,$\pm$\,500\,K, and ``high Teff'' – stars with \teff\,$>$ 6277\,K. The average errors on \teff\ are 38, 24, and 46 \emph{K} for cool, Sun-like, and hot star groups, respectively. The average errors in $\log{g}$ are 0.07, 0.03, and 0.05\,dex, in \emph{$\xi{}_{\mathrm{t}}$} - 0.09, 0.04, and 0.08\,km\,s$^{-1}$, and in [Fe/H] - 0.03, 0.02, 0.03\,dex for the three groups, respectively. We can see that for the three groups of stars, the dominant error is due to the uncertainty on continuum position, followed by the errors on \teff\ and \logg. On the other hand, carbon abundances are hardly sensitive to changes in [Fe/H] and $\xi_{\mathrm{t}}$. Moreover, for cool stars, the errors on the stellar parameters are larger and thus they translate into larger errors on abundances, especially those related to the \teff\ and \logg\ uncertainties. The final errors are given by the quadratic sum of these individual errors. The abundances and corresponding errors are provided in electronic tables.\\

\textit{Comparison with carbon abundances from CH band}\\

In Fig. \ref{comp_CH_band}, we show the differences between carbon abundances derived from \ion{C}{I} in this work and those derived from the CH band in \citet{suarez-andres17} for stars in common. We note that the spectra used to derive abundances from the CH band are the same as those employed here, even though the stellar parameters are slightly different for 308 out of the 757 stars. The average differences and standard deviation (ours - theirs) are the following: 7\,$\pm$\,28\,K for \teff, -0.018\,$\pm$\,0.092\,dex for \logg, and 0.001\,$\pm$\,0.015\,dex for [Fe/H]. We checked whether the observed differences in carbon abundances are due to the slight differences between stellar parameters, but when plotting the difference in carbon as a function of the difference in parameters, we could not see any trend, probably because the differences in parameters are small. However, the difference in abundances seem to depend on \teff\ and [Fe/H], as observed in Fig. \ref{comp_CH_band}. We find that the largest differences between both carbon indicators are found for cooler stars and less metallic stars, for which the atomic carbon lines provide a larger abundance. We note here the difficulties for measuring the CH band abundances in cool stars due to non-negligible blends in the region \citep[e.g.][]{pavlenko19}. On the other hand, the abundances from CH band tend to be higher than those from \ion{C}{I} lines for more metal-rich stars. This probably explains why we find a tendency for [C/Fe] ratios to increase as [Fe/H] decreases; whereas the work by \citet{suarez-andres17} reports a flatter trend the with metallicity (see Sect. \ref{sec:CFe_pop}). The carbon abundances from the APOGEE DR14 \citep[see Fig. 5 in][]{jonsson18} also show differences between the values derived from atomic lines (with a slight increasing trend of [C/Fe] towards lower metallicities) and a flatter trend of [C/Fe] across all the metallicity ranges for abundances obtained from carbon molecular lines. Nevertheless, a comparison between the above-mentioned studies is not straightforward since the APOGEE sample is mainly composed of red giants that suffer variations in their carbon content as they evolve in the Red Giant Branch.

\begin{figure}
\centering
\includegraphics[width=9.0cm]{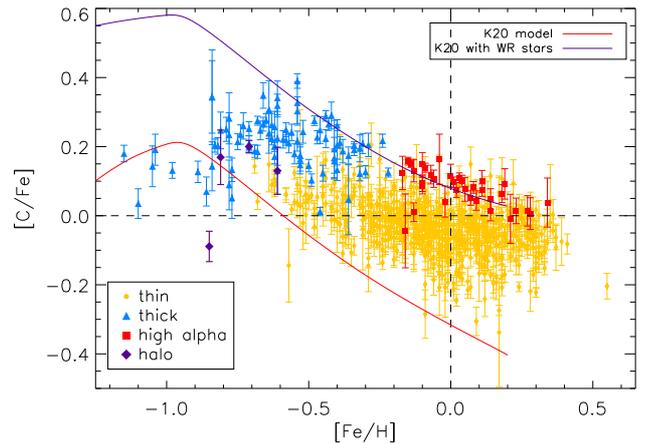}
\caption{[C/Fe] ratios as a function of [Fe/H] for the full sample including individual errors. The different stellar populations are depicted with different colours and symbols as explained in the legend. Models of carbon GCE from \citet{kobayashi20} are overplotted with a red line. Updated models considering the carbon production by Wolf-Rayet stars are plotted with a purple line (Kobayashi in prep.).} 
\label{CFe_pop}
\end{figure}

\begin{figure}
\centering
\includegraphics[width=8.7cm]{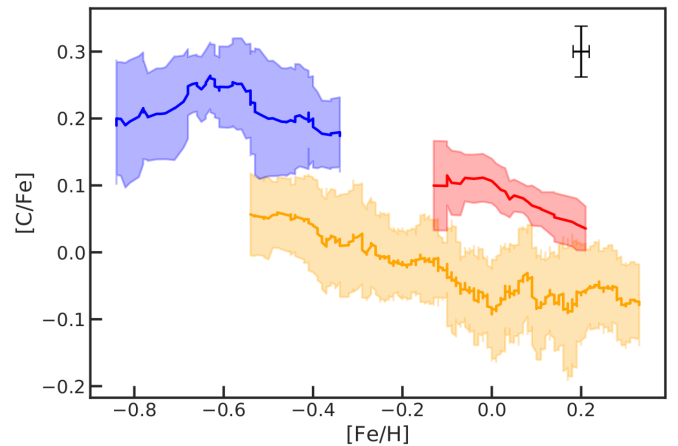}
\caption{Running mean of [C/Fe] as a function of [Fe/H]. The running means were calculated using windows of 30, 15, and 10 for thin disk (yellow line), thick disk (blue line) and the h$\alpha$mr (red line), respectively. These numbers reflect the sizes of each sample. The 1$\sigma$ uncertainty in the moving mean is also shown as a shaded region of the corresponding colour.} 
\label{CFe_pop_bins}
\end{figure}

\section{Carbon abundance ratios for different stellar populations} \label{sec:CFe_pop}

In this section, we discuss the behaviour of several carbon abundance ratios as a function of [Fe/H] and age for different stellar populations, as well as its implications for the nucleosynthesis production sites.

\subsection{[C/Fe] vs [Fe/H]}

In Fig. \ref{CFe_pop}, we show the distribution of [C/Fe] ratios as a function of metallicity for the different stellar populations. In total, we have 634 stars in the thin disk, 85 in the thick disk, 4 in the halo and 34 stars belong to the group of high-$\alpha$ metal-rich stars (hereafter, \textit{h$\alpha$mr}). The separation between those populations is based on a purely chemical criteria, except for the halo stars \citep[as explained in][and DM17]{adibekyan11}. At a first glance, the behaviour of [C/Fe] resembles that of some $\alpha$ elements such as Mg, Si, or Ti \citep[see][]{adibekyan12} or heavier elements such as Zn or Eu (see DM17), with a quite clear separation between the thin- and thick-disk stars. Interestingly, the \textit{h$\alpha$mr} stars also display higher abundances than thin-disk stars at the same metallicity. This separation between populations is better appreciated when plotting the running mean of [C/Fe] (see Fig. \ref{CFe_pop_bins}). Except for the recent paper by \citet{franchini20} that was based on the large data sample from the Gaia-ESO survey, we are not aware of any other study with a large sample showing so clear separation between the thick and thin disk as reported here. In addition, we also find a separation in carbon abundances between the thin disk and \textit{h$\alpha$mr} stars, clear up to [Fe/H]\,$\sim$\,0.2\,\,dex, since at higher metallicities there are not many \textit{h$\alpha$mr} stars with determined carbon abundances. We note that if the \textit{h$\alpha$mr} is not a separate population, but instead just the metal-rich tail of the thick disk \citep[as proposed by e.g.][]{bensby14}, then this result would mean that the thin-thick disk separation is clear up to super-solar metallicities. Finally, we find that [C/Fe] are below zero for thin-disk stars at solar metallicity, as also reported in previous works \citep[e.g.][]{franchini20,botelho20}, pointing to a possible carbon-rich nature of the Sun. However, this behaviour was not found in the works of \citet{jonay10,jonay13}, which were based on a subsample of stars with very high quality spectra either in the solar analogues or the 'hot analogues' parameter space. We checked whether by restricting our sample in the same way as the above-mentioned works the behaviour of [C/Fe] would be different at solar metallicity and we still find that thin-disk stars have an average [C/Fe] below solar as shown in Fig. \ref{CFe_pop_bins}. Therefore, differences may arise from the different set of atomic data, stellar parameters, or version of the radiative transfer code used in \citep{jonay10,jonay13}, despite employing the same spectra.

\begin{figure}
\centering
\includegraphics[width=9.0cm]{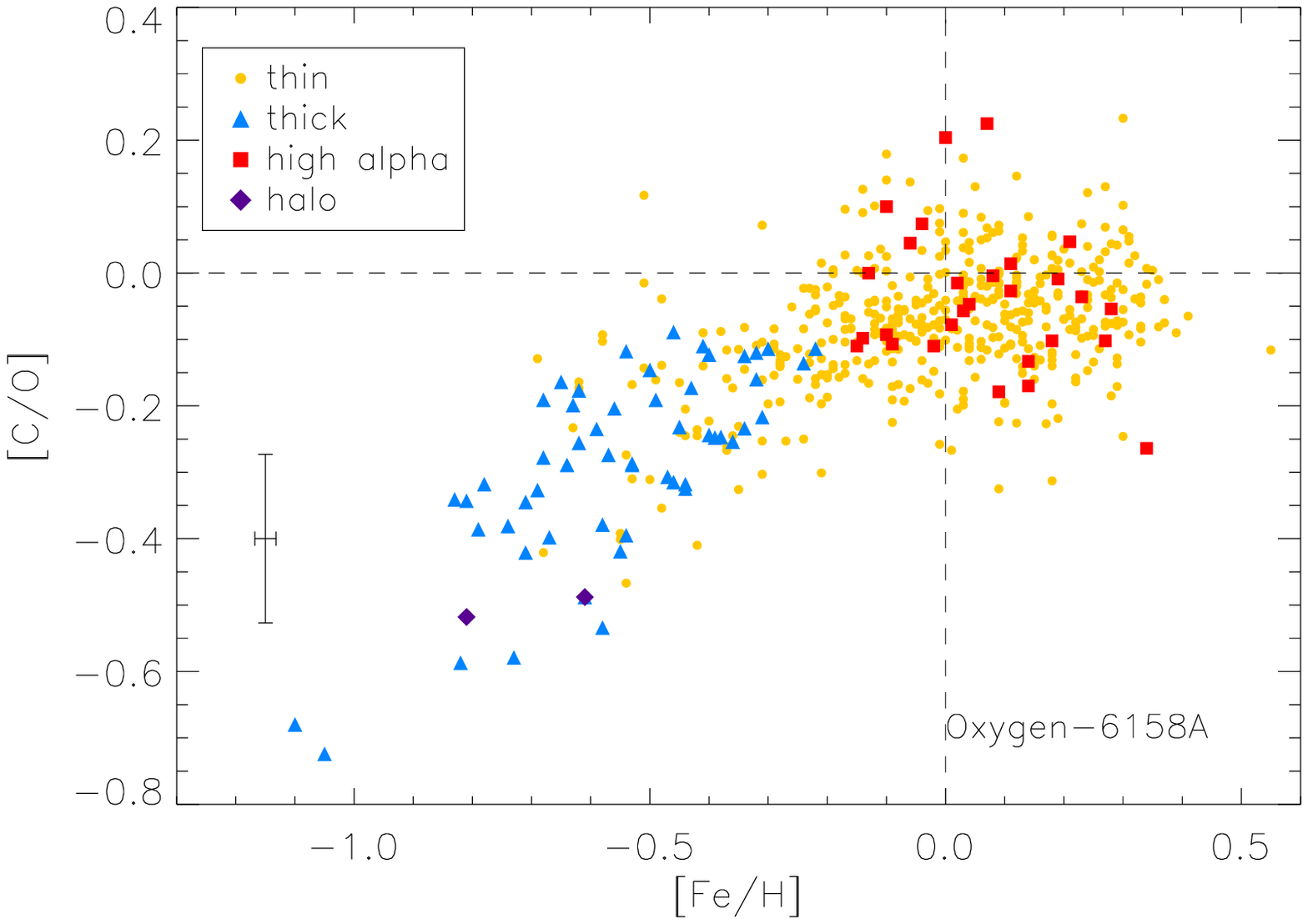}
\includegraphics[width=9.0cm]{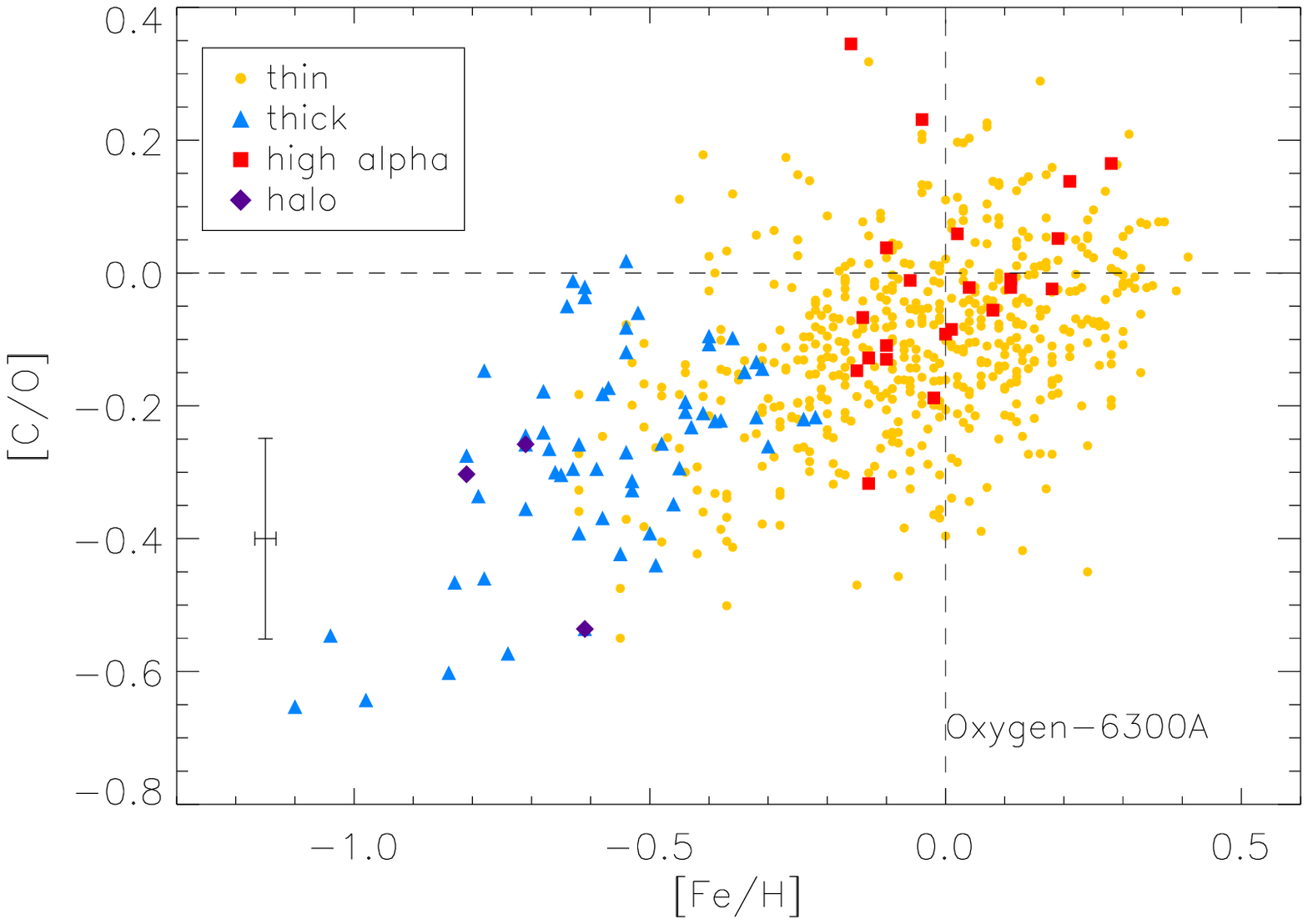}
\caption{[C/O] ratios as a function of metallicity using the oxygen indicators at 6158\AA{} or 6300\AA{}.} 
\label{CO_galaxy}
\end{figure}

Most of the previous studies also found increasing [C/Fe] ratios with decreasing [Fe/H] \citep[e.g.][]{takeda05_carbon,reddy06,delgado10,jonay10,jonay13,nissen14,buder19,franchini20,stonkute20} down to [Fe/H]\,$\sim$\,-0.8\,\,dex where the [C/Fe] flattens and then decreases for low-$\alpha$ halo stars \citep[e.g.][]{nissen14}. Despite that fact that we do not have many stars at low metallicities, we find a clear decrease in abundances for [Fe/H]\,$\lesssim$\,-0.7\,\,dex. Those mentioned works determined abundances of carbon with \ion{C}{I} optical lines, except for the work by \cite{reddy06}, which  also employs the lower excitation potential forbidden \ion{C}{I} line at 8727\AA,{} as well as the study by \citet{stonkute20}, which is only based on the C$_{2}$ Swan band head. On the other hand, the works by \cite{bensby06}, using the forbidden \ion{C}{I} line at 8727\AA{}, the study by \cite{suarez-andres17} based on the CH band, and the work by \citet{dasilva12} using both molecular and atomic carbon lines, report quite flat trends of [C/Fe] as [Fe/H] diminishes. Therefore, it seems that different carbon indicators can provide different trends as a function of metallicity, making it more difficult to constrain GCE models. This is especially critical at sub-solar metallicities, where NLTE effects are expected to have a greater impact \citep[e.g.][]{amarsi19}. Our [C/Fe] versus [Fe/H] trend points to a production of C by massive stars at lower metallicities, namely, at earlier times than the production of Fe by SNe Type Ia \citep[e.g.][]{carigi05}. At solar metallicity, there is a flattening of [C/Fe] for thin-disk stars which reflects the production of carbon by low- and intermediate-mass stars that release their products at a similar timescale as SNIa producing Fe. The most metallic star in our sample ([Fe/H]\,=\,0.55\,dex) points to a continuous decrease of [C/Fe] with increasing [Fe/H], as also reported by \citet[e.g.][]{amarsi19}. However, the lack of stars with determined [C/Fe] ratios in the metallicity range of 0.4-0.55\,dex prevents us from drawing a conclusion about the general trend at high metallicities. The maximum [C/Fe] found at [Fe/H]\,=\,-0.7\,dex matches the maximum production of carbon by asymptotic giant branch (AGB) stars \citep{andrews17}. In Fig. \ref{CFe_pop}, we overplot (with a red line) the GCE model from \citet{kobayashi20} that considers the production by carbon from SNe, super AGB stars, and AGB stars. This model shows systematically lower [C/Fe], as compared to our data but explains the trend of decreasing [C/Fe] ratios as [Fe/H] diminishes. The new set of models (purple line, Kobayashi in prep), including the contribution to carbon by rotating Wolf-Rayet stars, seem to better fit our data, although the abundance ratios are  overestimated in this case. 

\begin{figure}
\centering
\includegraphics[width=8.7cm]{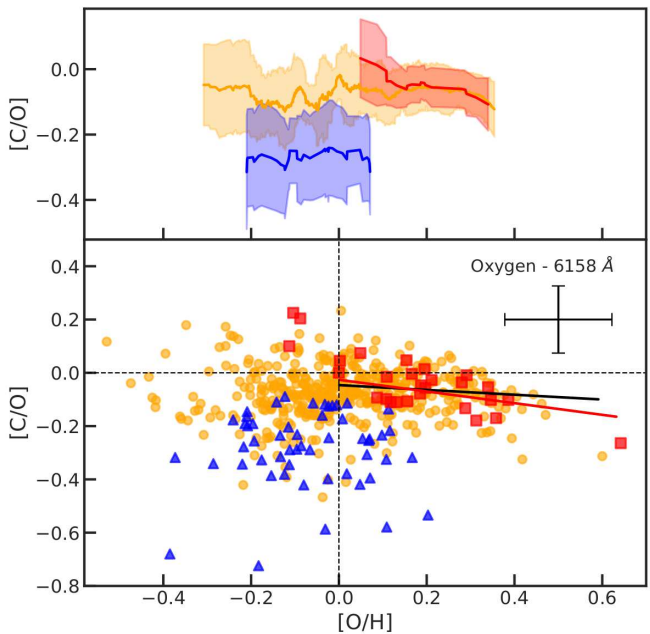}
\caption{[C/O] ratios as a function of [O/H] using the oxygen line at 6158\AA{} for the thin-disk, thick-disk. and h$\alpha$mr stars. The running means were calculated using windows of 30, 15, and 10 for thin-disk (yellow line), thick-disk (blue line) and the h$\alpha$mr (red line) stars, respectively. These numbers reflect the sizes of each sample. For these populations, the 1$\sigma$ uncertainty in the moving mean is also shown as a shaded region of the corresponding colour.} 
\label{CO_OH_galaxy_pop}
\end{figure}

\begin{figure}
\centering
\includegraphics[width=9.0cm]{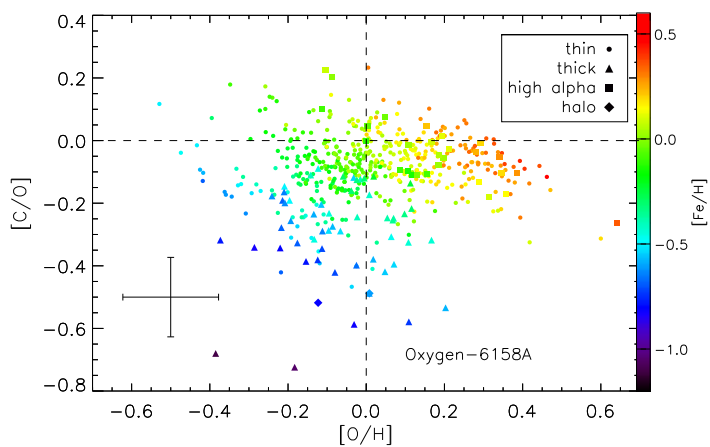}
\caption{[C/O] ratios as a function of [O/H], using the oxygen line at 6158\AA{} with a colour scale for metallicity.} 
\label{CO_OH_galaxy}
\end{figure}
\subsection{[C/O] vs [Fe/H]}

Another way of obtaining information about the nucleosynthesis processes involved in producing carbon is to compare it with other elements that are characterised by a well-known source of production, as in the case of oxygen. In Fig. \ref{CO_galaxy}, we show the variation of [C/O] as a function of [Fe/H], which serves as a first-order approximation to the evolution with time. To calculate the [C/O] ratios, two oxygen abundance indicators are used independently. At subsolar metallicities, the abundance ratios with both oxygen indicators are mostly negative and show an increasing trend towards higher metallicity. This is explained by the fact that oxygen is entirely produced by SNe Type II from massive progenitors, which started to release their yields at earlier ages in the Galaxy and, hence, at lower metallicities \citep[e.g.][]{woosley95}. The massive stars producing carbon at low metallicities might be less massive than those producing oxygen (i.e. having a longer life), explaining a delayed contribution of carbon, hence, the negative [C/O] ratios. Alternatively, this could be explained by increasing O/C yields for more massive progenitors of SNeII. Once metallicity starts to increase, low- and intermediate-mass stars release carbon and massive stars start to eject more carbon than oxygen \citep{carigi05}. The [C/O] ratio seems to have a constant rise towards higher metallicities when using the forbidden oxygen line. However, in the case when the \ion{O}{I} 6158\AA{} line is employed, we do observe that the maximum in [C/O] takes places close to solar metallicity to then become flat or decrease. This suggests that low-mass stars mostly contribute to carbon around solar metallicity, whereas at super-solar metallicities, massive stars produce carbon together with oxygen, thereby flattening or even decreasing the [C/O] ratio. This trend is in agreement with the metallicity dependent yields from \citet{carigi05}, which provide higher carbon as [Fe/H] increases from massive stars (i.e. also increasing the O production) but lower carbon from low and intermediate mass stars as [Fe/H] increases (i.e. less production of C). The turning point of increased relative production of carbon from massive stars takes place at A(O)\,$\sim$\,8.7\,dex \citep[see Fig. 2 of][]{carigi05} which equals to [O/H]$\sim$\,0.0\,dex. 
This observed behaviour of [C/O] is in contrast to the steady increase of [C/O] up to [Fe/H]\,$\sim$\,0.3\,dex found, for example, by \citet{franchini21}. Nevertheless, the general trend we find when using the $[\ion{O}{I}]$ 6300\AA{} line is similar to the reported by \citet{franchini21}, who use also that oxygen indicator. All thick-disk stars present negative [C/O] ratios and when using the oxygen line at 6158\AA{} thin-disk stars with [Fe/H]\,$\lesssim$\,--0.2 have [C/O]\,$<$\,0 as well. Thick-disk stars and low-metallicity thin-disk stars at the same metallicity have similar [C/O] ratios, meaning that the balance between different production sites for oxygen and carbon is the same among both populations, despite [C/Fe] and [O/Fe] being systematically higher for thick-disk stars at a given metallicity.

\subsection{[C/O] vs [O/H]}

In Fig. \ref{CO_OH_galaxy_pop}, we plot the [C/O] ratios dependence on [O/H] for the different stellar populations. This figure serves to evaluate the balance between the two different elements directly with the evolution of one of them, which in this case has a well-known, single production site. Here, we chose  to only show the ratios with the oxygen abundances from the 6158\AA{} indicator as it was shown to present less dispersion and be more trustworthy \citep{bertrandelis15}. Nevertheless, for completeness, we present the figure with the forbidden oxygen line in the appendix (Fig. \ref{CO_OH_galaxy_pop_6300}). Previous works in the literature have shown an increase of [C/O] ratios as [O/H] increases \citep[e.g.][]{bensby06,nissen14,amarsi19}, but our ratios present a quite large dispersion that together with the shorter range in [O/H] prevents us from seeing a clear behaviour. If we focus on the results with the 6158\AA{} line, the [C/O] ratios present a general flat trend (see the upper panel of Fig. \ref{CO_OH_galaxy_pop} with the running mean for each population), as [O/H] increases to then clearly decrease at [O/H]\,$\gtrsim$\,0\,dex, both for thin-disk and \textit{h$\alpha$mr} stars. To better evaluate the significance of this apparent trend, we applied a weighted least squares fit to the [C/O] values of thin disk and \textit{h$\alpha$mr} stars at [O/H]\,$\gtrsim$\,0\,dex. We then used the values of the slopes and the associated uncertainties to assess the significance of the fits. The p-values come from the F-statistics that tests the null hypothesis that the data can be modelled accurately by setting the regression coefficients to zero. The resulting p-values are 3.3\,e$^{-2}$ and 1.6\,e$^{-2}$ for the thin disk and \textit{h$\alpha$mr} stars, meaning that the correlations are significant. This trend is in agreement with the previously mentioned turning point of [C/O] at [O/H]$\sim$\,0.0\,dex \citep{carigi05}. However, this is in contrast with the flattening at super-solar [O/H] presented by \citet{nissen14} or the increasing trend found by \citet{franchini21} which might be caused by the use of different oxygen indicators. On the other hand, the [C/O] ratios for thick-disk stars present no trend but show a clear offset in its running mean with respect to the thin disk as also reported by \citet{amarsi19}. This separation between the thick and thin disk, also observed for other elements, supports the different formation episodes of both populations \citep{chiappini01,amarsi19}.

If we consider the [C/O] ratios with the forbidden line (Fig. \ref{CO_OH_galaxy_pop_6300}), the three populations of stars present an apparent decreasing [C/O] ratio for the full range of [O/H]. We also fitted the data of the three populations separately as explained before. The correlation of the slope found for thin-disk stars is not significant. However, the p-values for the slopes fitted for the thick disk and \textit{h$\alpha$mr} stars are 4.9\,e$^{-2}$ and 1.8\,e$^{-2}$, respectively, meaning that the correlation found is significant. \citet{franchini21} also found a decreasing [C/O] ratio with [O/H] by employing the same oxygen indicator but only for thick-disk stars. Therefore, it is hard to draw a conclusion about the behaviour of this ratio due to the variability observed in oxygen abundances. For example, the work by \cite{nissen14}, which uses both the forbidden line and the oxygen triplet at 7777\AA{} reports a slight increase of [C/O] with [O/H] both for thin- and thick-disk stars. This can be explained by the increased production of carbon from AGB stars as time passes or an increasing rate of carbon enrichment from metallicity-dependent winds from massive stars at later epochs \citep{amarsi19}. Nevertheless we note that the range in [O/H] of our work is much shorter than that of the previously mentioned works, a fact that could affect the general trend.

In order to better appreciate the behaviour of these two abundance ratios, we show in Fig. \ref{CO_OH_galaxy} the [C/O] versus [O/H] diagram with the stars grouped by metallicity (by colour scale). This figure shows that the [C/O] ratios decrease as [O/H] increases for stars of similar metallicity. We applied linear fits in different metallicity bins and we find the slopes of the [C/O] versus [O/H] are quite similar, with a mean value of -0.39. If we look at stars with similar [O/H], the increase of [C/O] is driven by a higher metallicity. This might be the reason why the [C/O] ratios of thick-disk stars are systematically lower than in thin-disk stars, as also reported in previous works \citep{bensby06,nissen14}. A similar behaviour is observed when using the forbidden line for oxygen (see Fig. \ref{CO_OH_galaxy_6300}), although in this case, the slopes of the [C/O] versus [O/H] relation in metallicity bins are steeper, with a mean value of -0.71.

\begin{figure}
\centering
\includegraphics[width=9.0cm]{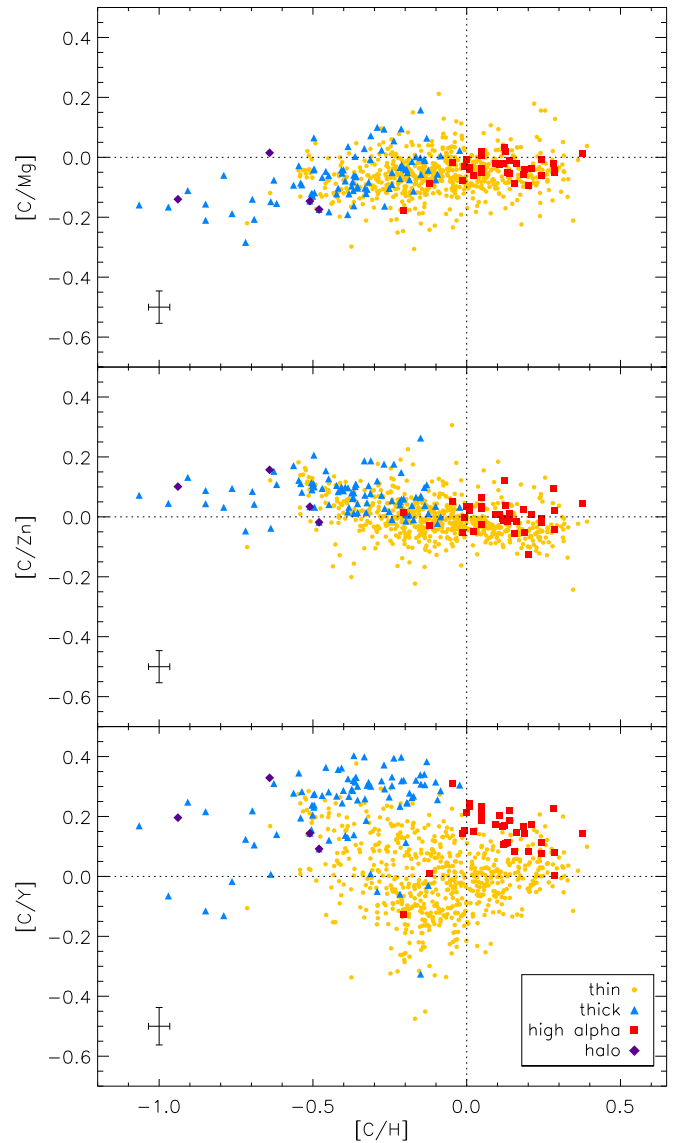}
\caption{[C/Mg], [C/Zn] and [C/Y] ratios as a function of metallicity.} 
\label{CH_heavy}
\end{figure}

\subsection{[C/Mg], [C/Zn], and [C/Y] versus [C/H]}

In this subsection, we compare the abundances of carbon with the abundances of elements produced in different sites. In the top panel of Fig. \ref{CH_heavy}, the behaviour of [C/Mg] versus [C/H] is shown. The abundance ratios for all the populations are well mixed and no clear offset is observed. There is an increasing trend of [C/Mg] as [C/H] increases up to solar carbon (more obvious for thick-disk stars) and most of the stars present negative [C/Mg] ratios. This suggests that at lower [C/H] ratios Mg might be produced by stars more massive than those producing C, thus enriching the medium with Mg at earlier times. Moreover, the SNe II yields of Mg are only mildly metallicity dependent but they increase with metallicity for C \citep{andrews17,carigi05}, which could explain the balance of [C/Mg] as [C/H] increases. 

[C/Zn] ratios show an decreasing trend with increasing [C/H] that continues at super-solar [C/H] ratios. At solar metallicities, Zn is believed to be produced in two ways, from SNe II for the most abundant isotope and from the weak \textit{s}-process (in massive stars) for the remaining isotopes \citep{bisterzo05}. To explain the overall trend of Zn the chemical evolution models need to also consider yields from HNe and SNe Ia \citep[e.g.]{kobayashi09}, but massive stars are the major contributors. In the thick disk, the [C/Zn] ratios are always above zero, whereas for the thin disk there is a higher dispersion. At super-solar [C/H], the trend is inverted and there is a higher production of Zn as compared to C.

The [C/Mg] and [C/Zn] ratios are quite similar between thick- and thin-disk stars of the same metallicity with both populations well mixed. However, this is not the case for the [C/Y] ratios which are clearly higher for thick-disk stars. This might be explained by the fact that in the thin disk, both carbon and \textit{s}-process elements such as Y are produced in a similar rate by low-mass stars (AGB stars) since thin-disk stars are younger and the yields from low-mass stars become available. On the other hand, for the older thick-disk stars the abundances of Y are lower due to delayed production by AGB stars, whereas carbon can be produced by more massive stars, hence having, as a consequence, high [C/Y] ratios.

\begin{figure}
\centering
\includegraphics[width=9.0cm]{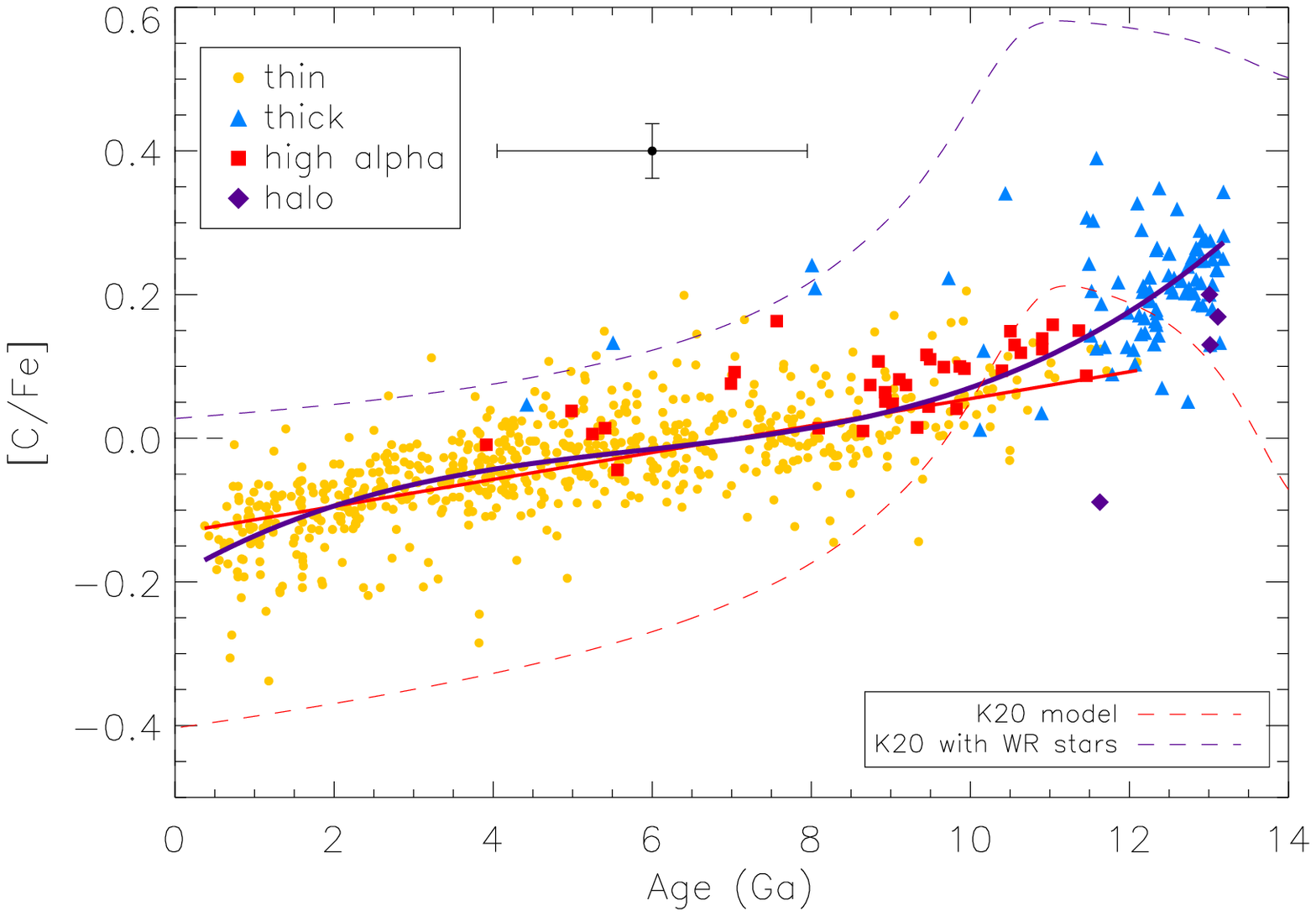}
\includegraphics[width=9.0cm]{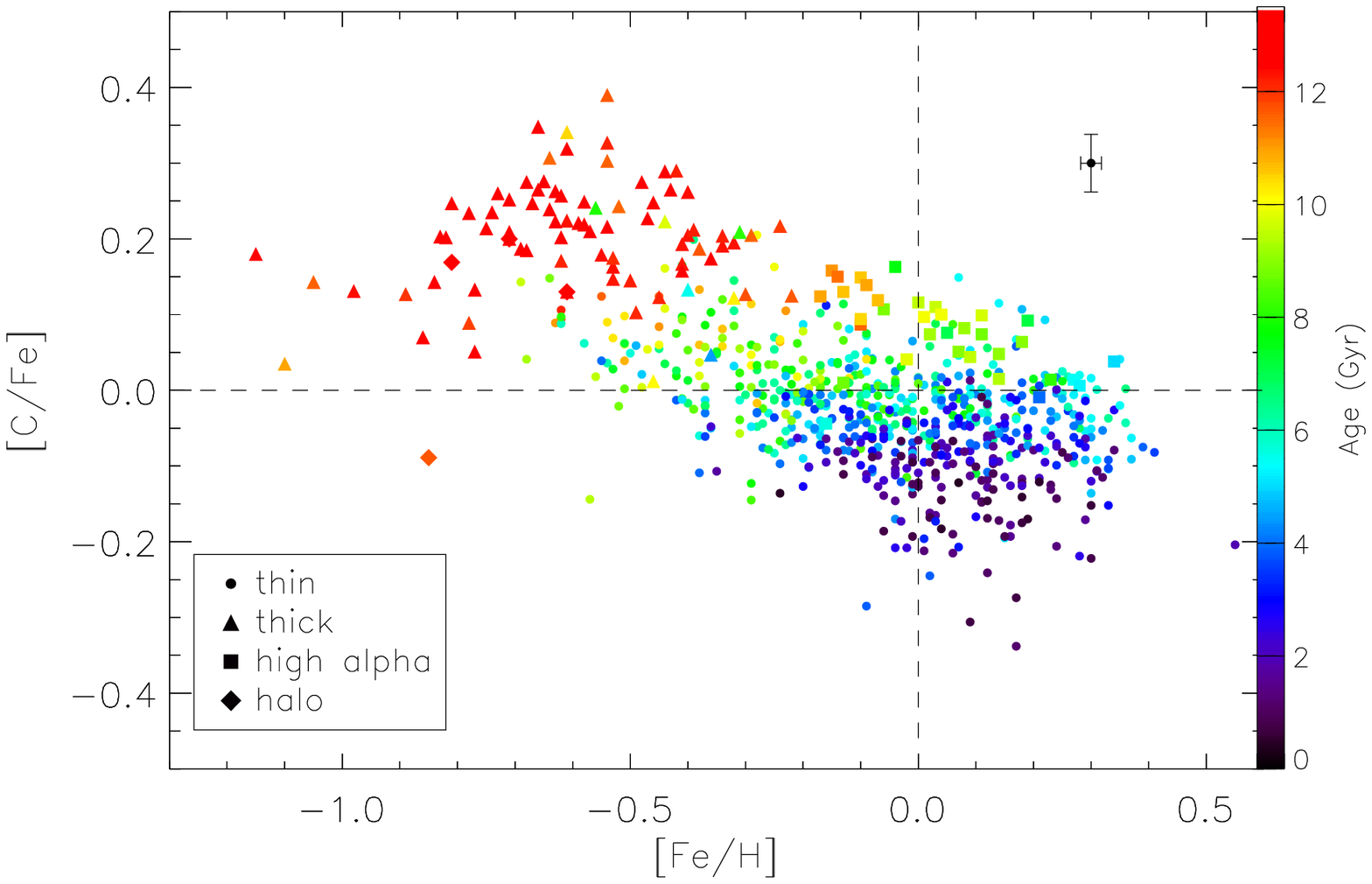}
\caption{[C/Fe] relation with age. Upper panel: [C/Fe] ratios as a function of age for the full sample. The red line is a weighted (by uncertainties) linear fit to the thin-disk stars to guide the eye on the general behavior of the trend. The thick purple line is a weighted fit with a third-order polynomial for all the stars. The dashed lines are the Kobayashi models also depicted in Fig. \ref{CFe_pop}. Lower panel: [C/Fe] as a function of [Fe/H] with ages in colour scale.} 
\label{CFe_age}
\end{figure}

\subsection{[C/Fe] evolution with age} \label{sec:CFe_age}

The exploration of the temporal evolution of abundances ratios can provide insightful information on production sites as well. For this purpose, we take advantage of the stellar ages that were already derived for the sample by using PARSEC isochrones \cite{bressan12}. We refer the reader to our previous work \citep{delgado19} for details. In the upper panel of Fig. \ref{CFe_age}, we present the variation of [C/Fe] ratios as a function of age. A simple linear fit to the thin-disk stars is added to visually appreciate the increasing trend for older ages, which has a slope of 0.019 \,dex/Ga: 

\begin{equation}
[C/Fe] = -0.1322 \pm 0.0018 + (0.0187 \pm 0.0003) \cdot Age
\label{eqCFe_age}
.\end{equation}

We note that we have not applied any constraints on the error on ages as we did for the analysis of abundance-age trends in our previous work \citep{delgado19}. If we restrict the sample to stars with errors in age below 1.5 Ga, we would obtain a [C/Fe]-age slope of 0.018\,dex/Ga. This slope is very similar to the one derived for Mg (0.021\,dex/Ga) for the same group of stars, below the steeper trend for oxygen (0.027\,dex/Ga) and above other $\alpha$ elements, such as Si (0.008\,dex/Ga) or TiII (0.016\,dex/Ga). This observed trend points to the potential of using carbon abundance ratios as a useful stellar age proxy \citep{jofre20} that is in line with previously proposed chemical clocks such as [Y/Mg], [Sr/Zn], or [TiII/Fe], among others \citep[e.g.][]{dasilva12,nissen15,delgado19}. We note, however, that care must be taken when dealing with stars with stellar parameters different than the Sun \citep{feltzing17,delgado19} or with stars in the inner disk of the Galaxy \citep{casali20}. In Fig. \ref{CFe_age}, we also add a fit with a third-order polynomial, which visually seems to better reproduce the data of the full sample: 

\begin{equation}
\begin{split}
 [C/Fe] = -0.1926 \pm 0.0041 + (0.0649 \pm 0.0025) \cdot Age, \\
 -(0.0089 \pm 0.0004) \cdot Age^{2} + (0.0005 \pm 0.00002) \cdot Age^{3} \end{split}
\label{eqCFe_age2}
.\end{equation}

This fit indicates that the production of carbon was faster for the oldest stars as found by \citet{franchini20,sharma21}, using the Gaia-ESO and GALAH surveys, respectively. The even steeper rise of [C/Fe] for thick-disk stars resembles that reported for other $\alpha$ elements, such as Mg or Ti in \citet{delgado19}. Furthermore, we find a steeper decrease of [C/Fe] for younger stars suggesting a reduction in the carbon production at recent times, which is opposite to what was reported by \cite{franchini20}, but more in agreement with \citet{sharma21}. The models of \citet{kobayashi20} reproduce the trend with age well, although they show a systematic shift with respect to our observations, as also seen in Fig. \ref{CFe_pop}.

In the lower panel of Fig. \ref{CFe_age}, the [C/Fe] vs [Fe/H] is shown again with the stellar ages on a colour scale. The pattern we observe here is very similar to other $\alpha$ elements \citep[see Figs. 5 and 8 from][]{delgado19}, although with somewhat higher dispersion\footnote{We note that the dispersion of [C/Fe] ratios is more than double than their average errors in all the metallicity bins and thus it seems of astrophysical errors and not caused by errors in carbon abundances}. When looking at stars of the same metallicity, there is a gradient of increasing [C/Fe] as the age increases that is prevalent at the full metallicity range.

\begin{figure}[h!]
\centering
\includegraphics[width=9.0cm]{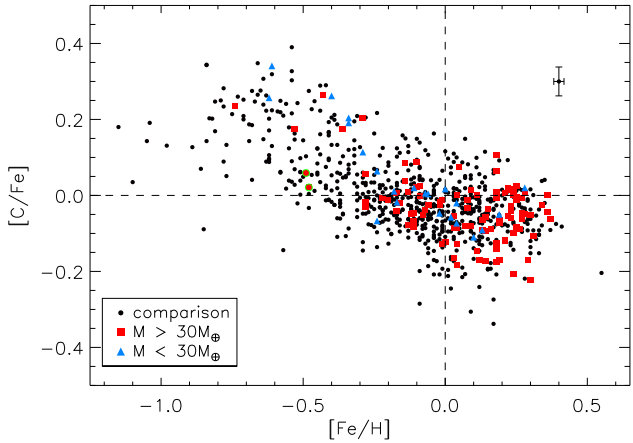}
\includegraphics[width=9.0cm]{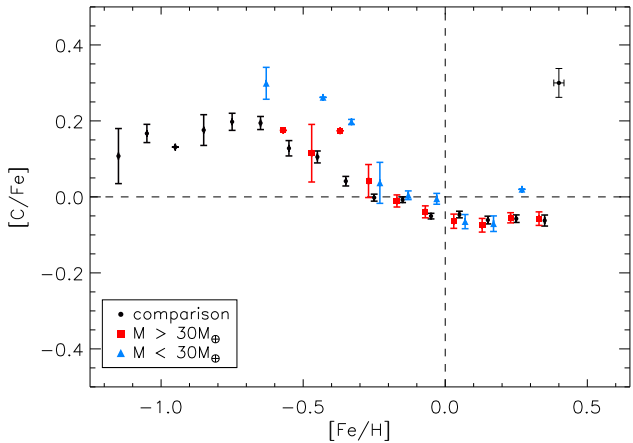}
\caption{[C/Fe] ratios in the populations of stars with and without planets. Upper panel: [C/Fe] as a function of metalliticy for stars with and without detected planets. Low-mass planets are depicted with blue triangles meanwhile stars with high-mass planets are shown with red squares. Black dots represent single stars. The two Jupiter-like hosts with a green circle around are HD\,171028 and HD\,190984. Bottom panel: Same as above but the mean abundances for each group of stars are shown in each metallicity bin of 0.1\,\,dex together with
the standard error of the mean.} 
\label{CFe_planets}
\end{figure}

\begin{figure}[h!]
\centering
\includegraphics[width=9.0cm]{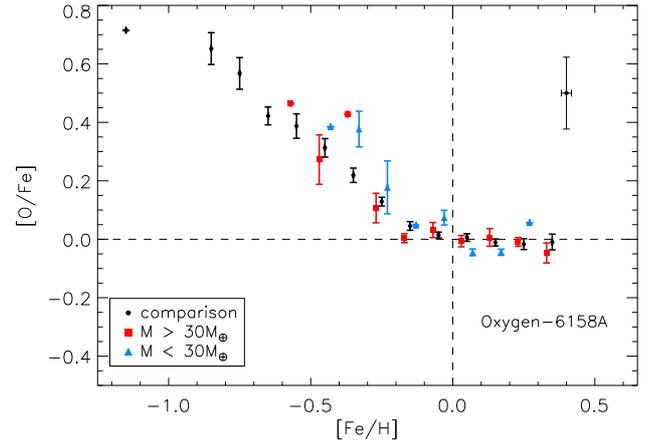}
\includegraphics[width=9.0cm]{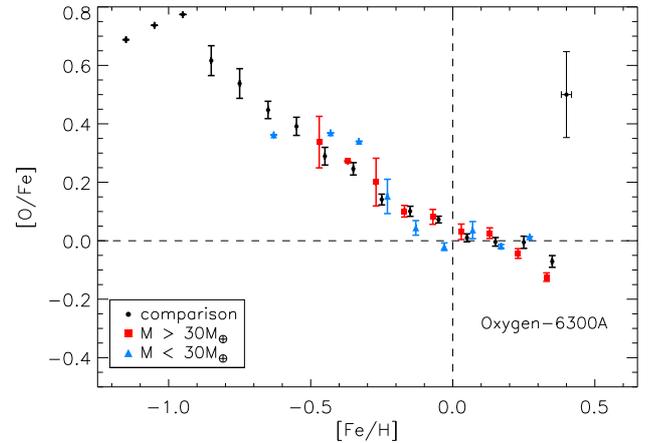}
\caption{[O/Fe] ratios in the populations of stars with and without planets. Upper panel: [O/Fe] as a function of metalliticy for stars with and without detected planets. The mean abundances for each group of stars are shown in each metallicity bin of 0.1\,\,dex, together with the standard error of the mean.} 
\label{OFe_planets}
\end{figure}

\section{[C/Fe] and [O/Fe] ratios for stars with and without detected planets}\label{sec:CFe_planets}

It is now well established that stars hosting giant planets are more metal-rich than stars without detected planets \citep[e.g.][]{gonzalez97,santos04,fischer05,sousa08}. On the other hand, more recent studies show that this correlation is not found for stars hosting less massive planets, those with masses as Neptune or lower \citep{sousa_harps4,buchhave12,everett13,buchhave15}. Furthermore, this correlation seems to also disappear when we consider very massive planets, namely, with masses above 4\,M$_{J}$ \citep{santos17,adibekyan19}.

The giant planet metallicity correlation supports the core-accretion scenario for the formation of planets \citep{pollack96,ida04,mordasini09}, in which it is assumed that planetesimals are formed by the condensation of heavy elements. The discovery of this correlation has led to an increased interest on the abundances of other elements in planet hosts \citep[e.g.][]{sadakane02,bodaghee03,beirao05,gilli06,ecuvillon06,bond06,robinson06,gonzalez07,takeda07,bond08,delgado10,jonay10,jonay13,kang11,brugamyer11,adibekyan15_MgSi,dasilva15,mishenina16,suarez-andres18}. Using the same sample of this study, \citet{adibekyan12_alpha} found that the [X/Fe] ratios of Mg, Al, Si, Sc, and Ti both for giant and low-mass planet hosts are systematically higher than those of stars without detected planets at low metallicities ([Fe/H]\,$\lesssim$ from --0.2 to 0.1\,\,dex depending on the element). Furthermore, this work confirmed the previous suggestion by \citet{haywood09} that planets form preferentially in the thick disk rather than in the thin disk at lower metallicities. A plausible explanation for this behaviour is that if the amount of iron is low, it needs to be compensated with other elements that are important for planet formation, such as Mg and Si, and these elements are more abundant in the thick disk \citet{adibekyan12_alpha,adibekyan12c}. In a subsequent work, \citet{delgado18} focused on heavier elements among the same sample of stars with and without planets, we found that planet hosts present higher abundances of Zn for [Fe/H]$<$--0.1\,\,dex, as a consequence of thick-disk stars having enhanced [Zn/Fe] ratios \citep{delgado17}. Moreover, \citep{delgado18} also found a statistically significant underabundance of Ba for low-mass planet hosts that had been previously suggested by \citet{mishenina16}. 

In the upper panel of Fig. \ref{CFe_planets}, we show the [C/Fe] abundance ratios for stars hosting at least one Jupiter-mass planet\footnote{M is defined as the minimum mass of the most massive planet in the planetary system of a star} (M\,$>$\,30\,M$_{\oplus}$), only Neptunians or super-Earths (M\,$<$\,30\,M$_{\oplus}$) and single stars. The number of stars with determined carbon abundances in each group are 101, 22, and 634, respectively. We can observe that for [Fe/H]\,$\gtrsim$\,--0.2, the three populations seem to be well mixed and there are no differences in carbon abundances as a consequence of planetary presence. However, for lower metallicities, it is obvious that planet hosts appear in the upper envelope of [C/Fe] ratios. This is more clearly appreciated in the lower panel of Fig. \ref{CFe_planets}, where the mean abundances in 0.1\,\,dex metallicity bins are depicted. Only in the bin of --0.5\,$<$\,[Fe/H]\,$<$\,--0.4\,\,dex, the stars with massive planets (red squares) have a similar average C as single stars. This is caused by the two slightly evolved stars, HD\,171028 and HD\,190984 (\logg\ values of 3.83\,\,dex and 3.92\,\,dex, respectively), which belong to the thin disk and do not present enhanced abundances of $\alpha$ abundances nor Zn \citep[see][]{adibekyan12_alpha,delgado18}. The reason for this C overabundance is mostly a result of the fact that planet hosts at low metallicities belong to the thick disk and, thus, they have higher abundances for several elements. However, if we compare those planet hosts in the thick disk at [Fe/H]\,$<$\,--0.2\,\,dex, with single stars only belonging to the thick disk, we still find that low-mass planet hosts have systematically higher [C/Fe]. In order to assess whether the visually appreciated differences between planet hosts and singles stars we applied a two-sided Anderson-Darling (AD) test (which is more sensitive to the tails of distributions than the Kolmogorov-Smirnov test). The AD test cannot reject the hypothesis that the samples of single stars and Jupiter hosts at [Fe/H]\,$<$\,--0.2\,\,dex come from the same population. On the other hand, The AD test statistic value for the comparison between single stars and Neptunian hosts has a significance value (p) of 5.8\,e$^{-2}$ which achieves the limit at which we can conclude that both populations are different. Indeed, the number of planet hosts in each bin is so small that this possible enhancement of C in the thick disk should be revised once more metal poor planet hosts are discovered. We note that in using a smaller sample, \citet{nissen14} also found indications of higher [C/Fe] ratio in planet hosts. However, these authors did not find a similar behaviour for oxygen and thus concluded that this carbon enhancement could not be caused by a higher volatile-to-refractory ratio in planet hosting stars, as was suggested in previous works \citep[e.g.][]{ramirez14}. Nevertheless, the finding of a deficiency of refractory elements in low-mass planet hosts has been widely discussed in the literature and other works do find both positive and negative T${_C}$ (condensation temperature) slopes \cite[e.g.][]{jonay10,jonay13} for terrestrial planet hosts. Moreover, the volatile-to-refractory ratios might be related to stellar age and Galactocentric radii, rather than to the presence of planets \citep{adibekyan14,adibekyan16}.

For comparison we also checked the behaviour of [O/Fe] ratios in the samples of stars with and without planets in different metallicity bins (see Fig. \ref{OFe_planets}). At a first glance, this looks similar to the trend for [C/Fe] (lower panel of Fig. \ref{CFe_planets}); that is, low-mass planet hosts show enhanced values of [O/Fe] in the lowest metallicity bins. However, the larger errors on oxygen abundances (and the different results for the two oxygen lines) made us refrain from setting a strong conclusion on this regard. We note that a higher content of oxygen in the bulk composition of these low-mass planets would imply that they have high water-mass fraction and lower core-mass fractions \citep{santos17} in comparison with planets orbiting stars in the thin disk.

\section{Fluctuations of C/O elemental ratios due to different factors}\label{sec:CO_variation}

In this section, we explore the C/O elemental\footnote{Also called C/O number ratios in the literature to distinguish them from the [C/O] ratios which are derived respect to the solar values.} ratios in our sample and the different factors that may produce large differences between ratios derived in different works.

The determination of the mineralogical ratios C/O and Mg/Si is useful in better constraining the interiors of rocky planets \citep[e.g.][]{bond10,dorn15}. Since planets form from the same material as their host stars it is generally assumed that rocky planets can share the ratios between C, O, Si, Mg, and Fe with their host stars. However, in a recent work, \citet{adibekyan21}  showed that despite there being a clear correlation between the iron mass fraction of the stars and the iron mass fraction of their planets (for rocky planets, as obtained from their mass and radius) this is not a one-to-one correlation as previously assumed. Therefore, this hypothesis might be taken as a first-order approximation when studying the Mg/Si ratio but the low condensation temperatures of the light elements carbon and oxygen make the planetary C/O ratio more dependent on the distance to the star where the planet was formed \citep[e.g.][]{thiabaud15,brewer17}. Therefore, carbon-rich planets can still be formed around oxygen-rich stars \citep[][]{madhu11,oberg11}. 

\begin{figure}
\centering
\includegraphics[width=9.0cm]{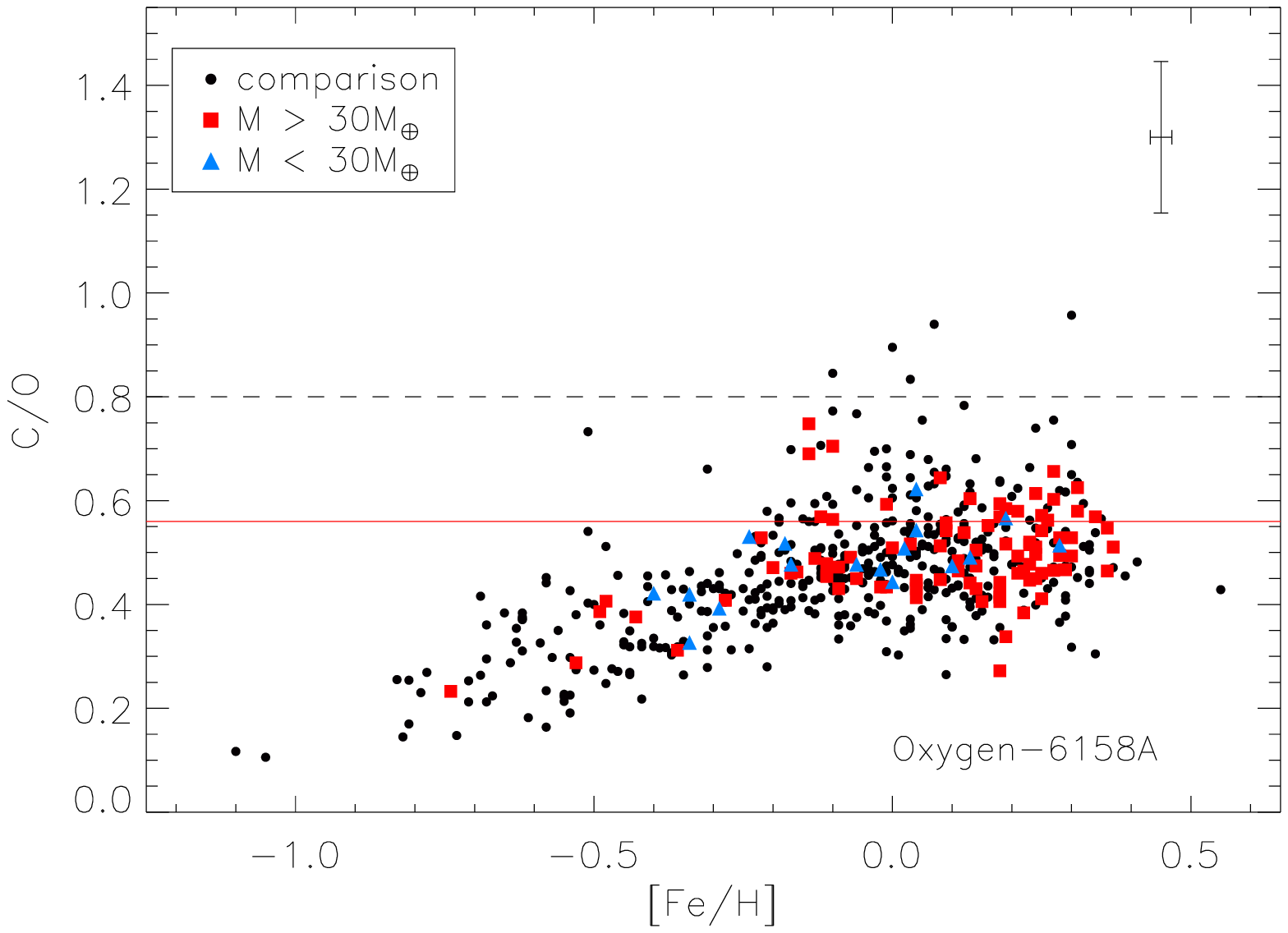}
\includegraphics[width=9.0cm]{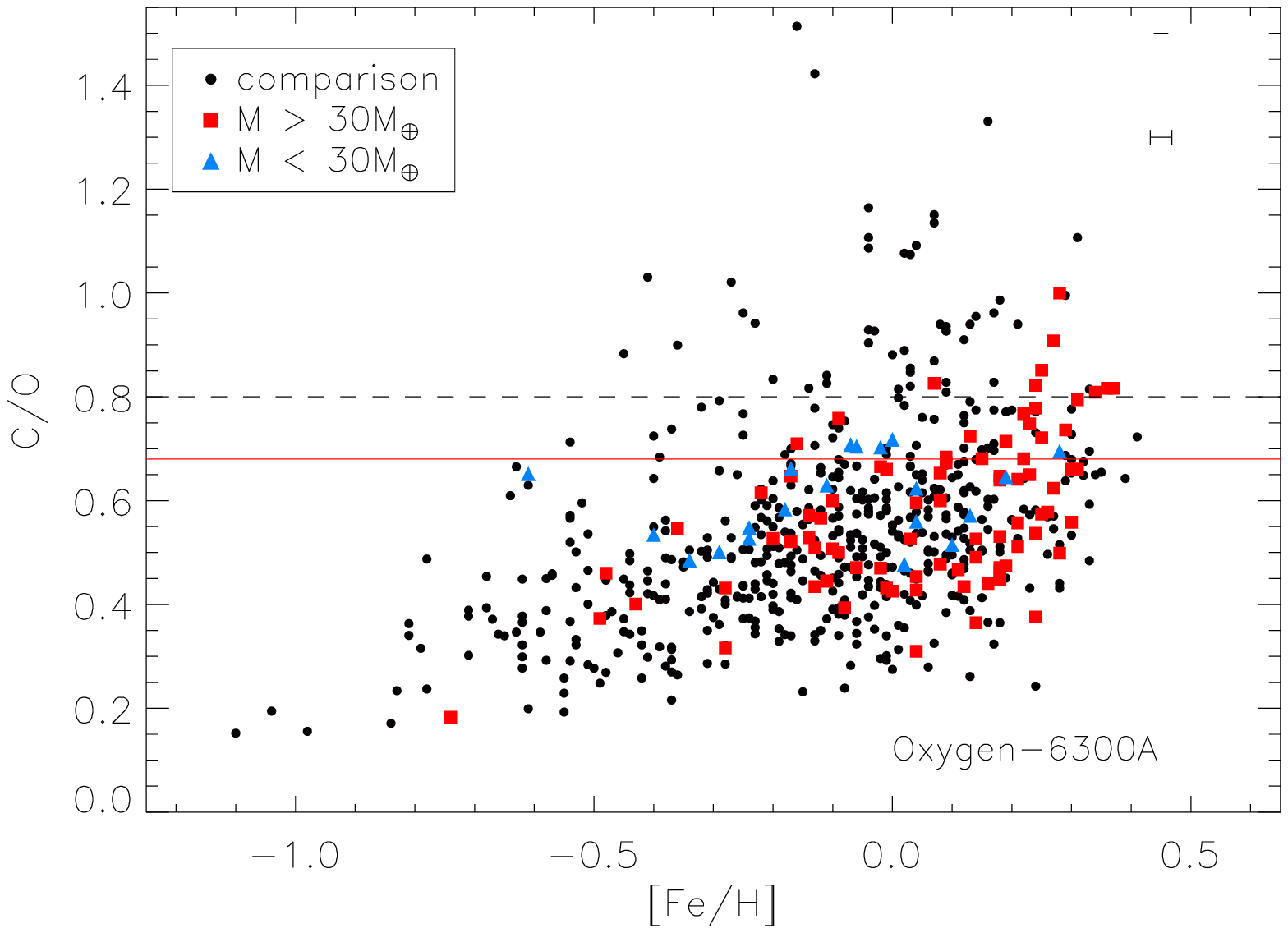}
\caption{C/O elemental ratios as a function of [Fe/H] for different oxygen indicators. The red continuous line marks the solar C/O value for each oxygen indicator. The black dashed line marks the limit of C/O values above which planets are considered as carbon-rich.} 
\label{CO_planets}
\end{figure}

In Fig. \ref{CO_planets}, the C/O values for stars with and without detected planets are shown. The elemental C/O ratios for the full sample increase as metallicity increases up to solar metallicity and then they flatten -- when using the oxygen 6158\AA{} line \citep[as found in e.g.][]{brewer16}. This trend is however increasing in all the metallicity range when calculating C/O ratios with the forbidden oxygen line \citep[as found in other works based on smaller samples e.g.][]{nissen13,teske14}, but as mentioned previously, we have greater trust  in the abundances from the oxygen 6158\AA{} line. We evaluated the C/O distributions (derived with the oxygen 6158\AA{} line) for the three groups of stars in the above mentioned figure and we find very similar average C/O ratios, 0.46\,$\pm$\,0.13, 0.49\,$\pm$\,0.09, 0.48\,$\pm$\,0.07, for the comparison sample, Jupiter-type or Neptunian-type companion samples, respectively, all of them below the solar value. However, most of the planet hosts present C/O ratios below 0.7, whereas the comparison sample has a small tail of stars with C/O ratios reaching to 1. Nevertheless, those higher C/O ratios are still compatible with a value close to 0.8 given the high errors associated with the elemental C/O ratios. The histogram with the C/O distribution is shown in the appendix (see Fig. \ref{CO_histogram}). To better compare these distributions we applied a two-sided Anderson-Darling (AD) test (which is more sensitive to the tails of distributions than the Kolmogorov-Smirnov test). The AD test statistic value for the comparison between single stars and Jupiter hosts has a significance value (p) of 1.6e$^{-3}$, suggesting that these distributions come form a different parent population. On the other hand, the AD test cannot reject the hypothesis that the samples of single stars and Neptunian hosts come from the same population. This dissimilarity of C/O ratios between Jupiter hosts and single stars can be inherent to the different metallicity distributions of both populations \citep[since C/O tends to increase with metallicity, see also][]{suarez-andres18}. We also check whether the C/O distributions of the Jupiter and Neptunian hosts are different, but the AD test provides a not significant p-value. A comparison of C/O ratios as a function of planet mass does not reveal any trend either. Therefore, we can conclude that the C/O ratios distributions of stars hosting planets of different masses are similar.

\subsection{C/O ratios dependence on oxygen indicators and solar reference}\label{sec:CO_solar}

In the literature, it is common to consider the solar C/O ratio as determined with solar abundances from \citet{asplund09}, with values of 0.55 (A(C)=8.43, A(O)=8.69). The elemental C/O ratios are calculated as follows:

\begin{equation}
C/O = N_{\rm C}/N_{\rm O} = 10^{A(C)}/10^{A(O)}
\label{eqCR}
\end{equation}

Based on this solar accepted ratio a classification of different internal compositions can be made in terms of C/O and Mg/Si elemental ratios \citep{bond10}. Here, we only focus on that which is related to the C/O ratios. For more information on the Mg/Si ratios obtained for this sample, we refer the reader to the previous works published by our group of authors \citep{suarez-andres18,adibekyan15_MgSi,delgado10}. Following the classification made by \cite{bond10} we could expect planets with C/O ratios higher than 0.8 to have Si mostly in the form of SiC because most of the oxygen is trapped in CO and there is hardly free oxygen to form silicates. On the other hand, in those systems with C/O ratios lower than 0.8, Si will be present in rock-forming minerals such as SiO$_{2}$. Those systems with C/O ratios higher than 0.8 have been referred as carbon-rich worlds \citep[e.g.][]{madhu12} and we do not expect to find many of them. However, recent observations of very young stars with ALMA radiotelescope have shown that disk C/O ratios higher than one are needed to explain the detected abundances of certain molecules in protoplanetary disks \citep{semenov18,facchini21}. Therefore, it is plausible to find planets with high C/O ratios but their host stars do not necessarily need to have such a large C/O ratio \citep[e.g.][]{teske14,brewer17}. Nevertheless, the difficulty in deriving precise oxygen abundances from weak lines (and its large associated errors) can lead to unrealistic high values of stellar C/O. In Fig. \ref{CO_planets}, we can see how the ratios derived with the $[\ion{O}{I}]$ line at 6300$\AA{}$ are systematically higher as compared with the ratios derived with the line at 6158$\AA{}$. In fact, when using the  forbidden line of oxygen, we find that almost 10\% of the stars (either in the high-mass planet hosts group or single stars group) have C/O ratios that are higher than 0.8. On the other hand, this percentage goes down to 1\% when considering the oxygen indicator at 6158$\AA{}$. In our first work focussing on C/O ratios for a subsample of the stars analysed here \citep{delgado10}, we only used the  forbidden line of oxygen to determine C/O ratios. This resulted in a significant number of stars having very high values of C/O, especially for cooler stars since carbon abundances present a slight enhancement towards lower \teff\ (see Fig. \ref{CFe_teff}); meanwhile, oxygen abundances decrease slightly for cooler stars \citep[see Fig. 5 of][]{bertrandelis15}. We also checked the behaviour of C/O vs \teff\ and found that it shows a slight increasing trend towards lower effective temperatures for C/O ratios obtained with the oxygen forbidden line, whereas it is flat for the other line. Therefore, it seems that using the C/O ratios derived with the oxygen line at 6158$\AA{}$ is a more reliable approach and indeed the percentage of stars with determined high C/O ratios is in line with previous studies on solar-type stars or sub-dwarfs \citep[see e.g.][]{fortney12}. Moreover, the fact of using abundances determined from lines with high excitation potential (as is the case for $[\ion{O}{I}]$ line at 6158$\AA{}$ and the optical $[\ion{C}{I}]$ lines) can provide abundance ratios that are less prone to systematic effects such as abundance dependence on \teff\ \citep[e.g.][]{nissen13}.

A possible way of overcoming this inhomogeneity on the determination of stellar C/O ratios due to different oxygen indicators is to consider the different solar reference values used in a given study \citep{fortney12,nissen13,teske14}. In Fig. \ref{CO_planets} our solar C/O values is depicted with a red continuous line, being 0.56 for the line at 6158$\AA{}$ and 0.68 for the line at 6300$\AA{}$. Indeed, our solar C/O ratio for the oxygen forbidden line is significantly larger than the standard value of 0.55 and it explains why the C/O ratios for all the stars are shifted towards higher values. However, when considering the [C/O] ratios (which are calculated respect to solar values) rather than the number C/O ratio (which is based in the absolute abundances) the number of stars with high values is more similar for both oxygen lines, as shown in Fig. \ref{CO_galaxy}. Therefore, when using the abundance ratios provided in this work, we recommend  scaling them to the desired solar reference (C/O$_\odot$), considering our ratios relative to the Sun, [C/O], defined as follows:

\begin{equation}
[C/O] = [C/H] - [O/H],
\label{eqCR2}
\end{equation}

where [C/H] = A(C)-A(C)$_\odot$ and [O/H] = A(O)-A(O)$_\odot$. By using Eqs. \ref{eqCR} and \ref{eqCR2} we can determine the scaled C/O ratios as:

\begin{equation}
C/O_{scaled} = 10^{[C/O]} * C/O_{\odot}
\label{eqCR3}
.\end{equation}

In this way, the obtained C/O ratios will be less dependent on systematics that are inherent to specific methods to determine stellar abundances -- such as the employed atomic data, radiative transfer code, or stellar atmosphere models (see next subsection) as well as comparisons between different works -- can be made in a more straightforward way.

\subsection{C/O ratios dependence on model atmospheres}

\begin{figure}
\centering
\includegraphics[width=9.0cm]{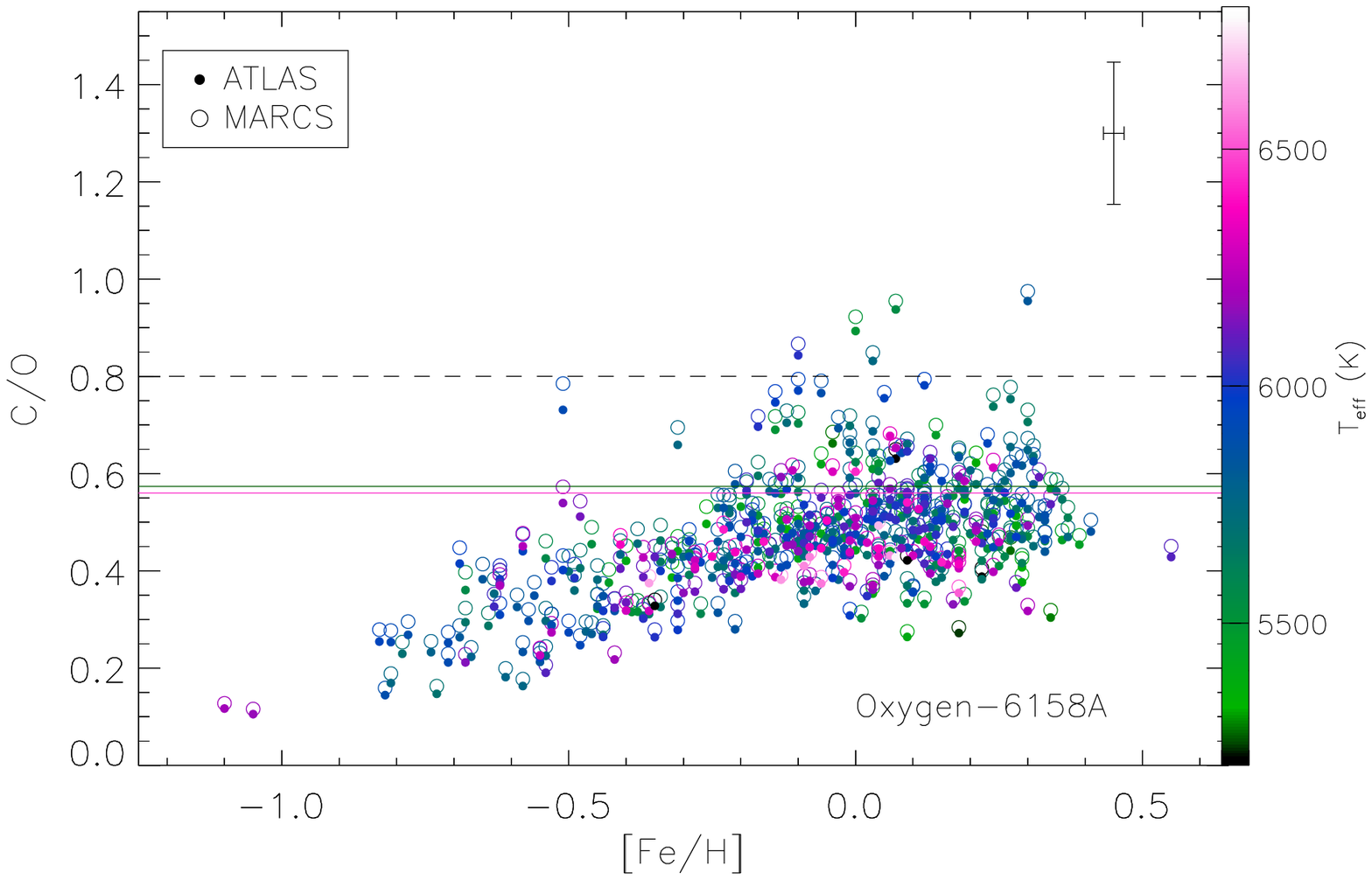}
\includegraphics[width=9.0cm]{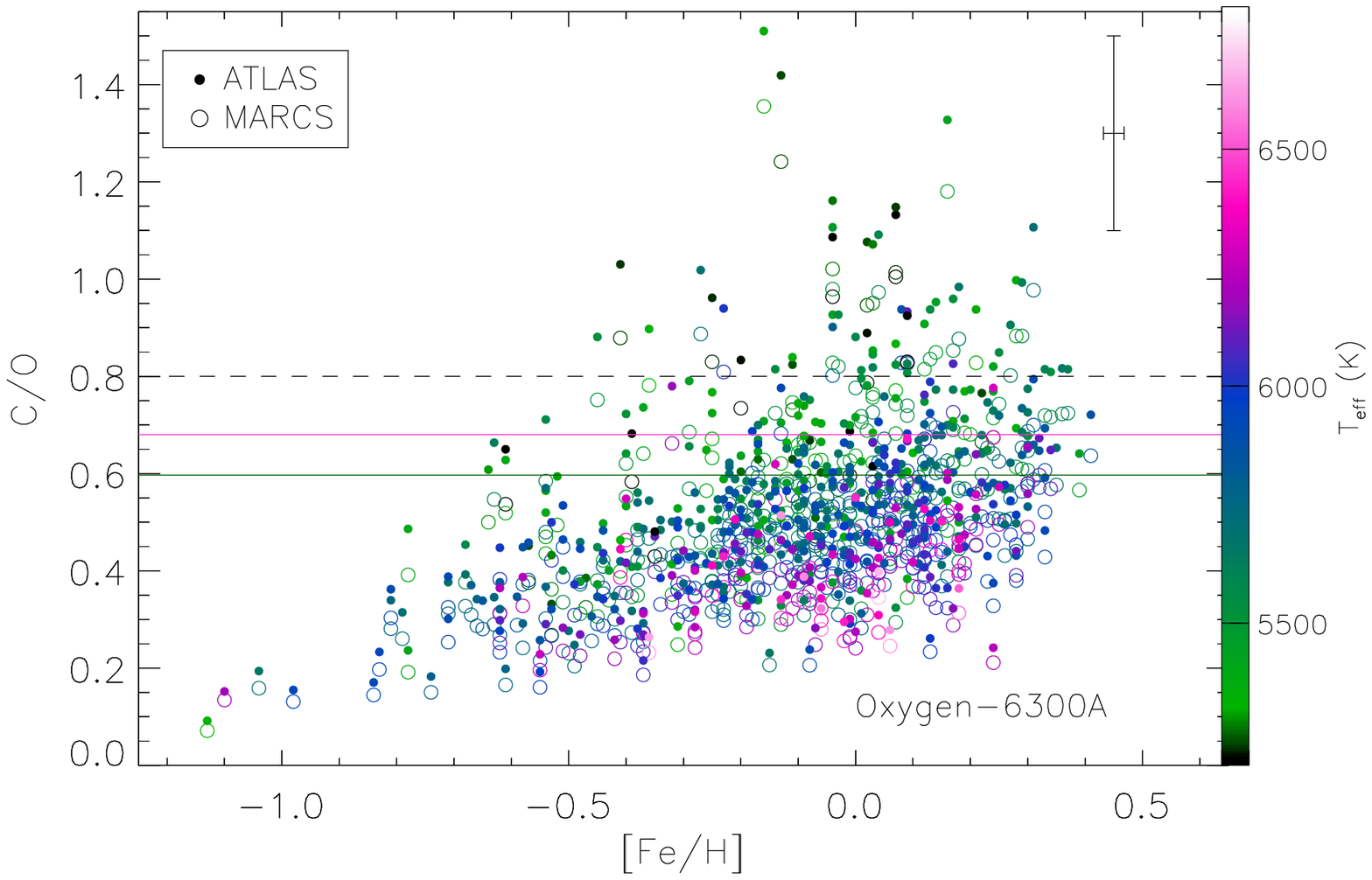}
\caption{C/O elemental ratios as a function of [Fe/H] for both oxygen indicators using abundances derived with ATLAS models (filled circles) or MARCS models (open circles). The \teff\ of each star is depicted with a color scale. The magenta and green continuous lines marks the solar C/O value for ATLAS and MARCS models, respectively. The black dashed line marks the limit of C/O values above which planets are considered as carbon-rich.}
\label{CO_models}
\end{figure}

In this subsection, we explore the possible influence on C/O ratios when using different kind of models atmospheres. As such we compare our results, obtained with ATLAS model atmospheres with those derived by using MARCS models \cite{gustafsson08}. These two sets of atmospheres are widely used in solar-type studies. In Fig. \ref{CO_models}, we show the C/O ratios obtained with the 6158$\AA{}$ oxygen line (upper panel) and the 6300$\AA{}$ oxygen line (lower panel). For the first oxygen indicator, the differences are quite small, with a median value of 0.017 (MARCS models provide slightly higher C/O ratios than ATLAS models) and a maximum difference of 0.054. On the other hand, the differences for the forbidden line are higher (the median value is 0.065, with ATLAS models providing larger values) and can attain values as large as 0.177 for the coolest stars. The differences in C/O ratios are also visible for the Sun (depicted by continuous magenta and green lines). Carbon abundances for the Sun using MARCS models are $\sim$\,0.1\,dex lower than with ATLAS models for both carbon lines. In the case of oxygen, the abundance of the 6158$\AA{}$ line is $\sim$\,0.12\,dex lower using MARCS as well (hence, balancing the C/O ratios), whereas it is only $\sim$\,0.05\,dex lower in the case of the 6300$\AA{}$ line, leading to a larger difference. Since C/O number ratios are dependent on the model atmospheres, it is therefore advisable to first determine [C/O] ratios using solar abundances derived with the same model atmospheres and then scale such [C/O] ratios with the desired solar C/O number ratio, as indicated earlier in this paper. Figure. \ref{CO_models} also highlights the C/O dependence on \teff\ when using the forbidden oxygen line, since the abundances of this line have a slight decreasing tendency towards cooler temperatures \cite[see Figure 5 in][]{bertrandelis15}, meanwhile carbon abundances slightly increase for cooler stars (as shown in Sect. \ref{sec:abundances}). On the other hand, the abundances from the 6158$\AA{}$ oxygen line have a slight tendency to increase as \teff\ diminishes (in the same way as carbon abundances), cancelling out the \teff\ effect in C/O ratios.

\section{Summary}\label{sec:summary}

In this work, we present carbon abundances obtained from optical atomic lines for a sample of 757 stars within the HARPS-GTO planet search sample. This paper belongs to a series of works studying the chemical abundances in the aforementioned sample with a twofold objective: evaluate the abundance ratios in the context of the chemical evolution of the Galaxy and explore the elements that show differences in the populations with and without discovered planets.
The results of this work can be summarised as follows:

\begin{itemize}

 \item We find a clear separation of [C/Fe] ratios between the thin- and thick-disk populations in a similar way as has been found for $\alpha$ elements. We show for the first time that high-$\alpha$ metal-rich stars also have enhanced [C/Fe] ratios when compared to thin-disk stars at the same metallicity. This trend is observed up to [Fe/H] $\sim$ 0.2, suggesting that [C/Fe] ratios for low- and high-$\alpha$ populations merge at higher metallicities in the same way that was found for $\alpha$ elements \citep{adibekyan11} and based upon this same sample of stars. The models by \citet{kobayashi20} can aptly reproduce the general trend of [C/Fe] vs [Fe/H], but they do show a systematic offset with our data.
 
 \item We explore the behaviour of [C/O] ratios as a function of [Fe/H] and [O/H]. The [C/O] ratios are similar for stars with similar metallicities in the thick- and thin-disk, thus, the balance between the production of these elements is similar in both populations. However, when [C/O] is compared with [O/H], the thick disk is shifted towards lower [C/O] ratios as compared to the thin disk. The [C/O] ratios present a quite flat trend in the different populations with a more clear decrease at super-solar [O/H] ratios in agreement with the models by \citet{carigi05} that predict a decrease of carbon production from low and intermediate mass stars for high metallicities. This behaviour is different as reported in other works in the literature and it might be caused by the shorter [O/H] range of our sample. The [C/O] decrease with increasing [O/H] is more obvious when we look at groups of similar metallicity. It is worth mentioning that the trends of [C/O] ratios are dependent on the oxygen lines used to derive them and so this might explain the discrepancies among different works in the literature. Due to the lower dispersion of oxygen abundances derived with the 6158$\AA{}$ line, we have greater trust in the results obtained from it.
 
 \item The [C/Fe] ratios present an increasing trend with stellar age with a linear slope of 0.019 dex/Ga, very similar to the slope found for [Mg/Fe] but less steep that the slope for [O/Fe], which probes the potential of [C/Fe] ratios as an age proxy. The general trend of [C/Fe] is better fitted by a polynomial since thick-disk stars present a steeper slope supporting the fact that the production of carbon was faster for the oldest stars. We also find indications of a steeper decrease of [C/Fe] for the youngest stars which indicates a possible reduction of carbon production in recent times.
 
 \item We find that low-mass planet hosts at [Fe/H]\,$<$\,--0.2\,dex have higher [C/Fe] ratios as compared to single stars at similar metallicities in the same way as found for $\alpha$ elements by \citet{adibekyan12_alpha}. With the current data, we cannot distinguish whether this enhancement in carbon is a consequence of planet hosts belonging to the thick disk or whether it is a factor favouring the formation of planets as seems to be the need of higher abundances of planet building materials such as Fe, Mg and Si. Nevertheless we note that this difference between [C/Fe] distributions is at the limit of being considered statistically significant and thus this trend should be revised in the future with a larger sample of low-mass planet-hosts.
 
 \item The C/O ratios derived with the oxygen 6158$\AA{}$ line have values below 0.8 for the vast majority of the stars whereas the oxygen forbidden line provides systematic higher C/O ratios. We compare the distributions of C/O ratios between stars hosting Neptunian or Jupiter type planets and single stars. We only find a statistically significant difference between the stars hosting Jupiters and single stars, although we note that this might be caused by the inherent different [Fe/H] distributions of such populations since C/O increases with metallicity.
 
 \item Differently from abundance ratios relative to the Sun such as [C/Fe], [C/O], or [O/H] (to name a few), the C/O elemental ratios are prone to large systematic effects. These may be produced through the application of different abundance line indicators, solar reference abundances, and model atmospheres. Therefore, C/O ratios for a same star can largely vary between different studies and it is not recommended to mix C/O ratios from different sources. The best solution to overcome this issue is to use the available [C/O] ratios (relative to the solar reference used in each work) and scale them to the desired solar C/O elemental ratio.

\end{itemize}

\begin{acknowledgements}
E.D.M., V.A., N.C.S. and S.G.S. acknowledge the support from Funda\c{c}\~ao para a Ci\^encia e a Tecnologia (FCT) through national funds
and from FEDER through COMPETE2020 by the following grants UIDB/04434/2020 \& UIDP/04434/2020 \& PTDC/FIS-AST/28953/2017, POCI-01-0145-FEDER-028953 \& PTDC/FIS-AST/32113/2017, POCI-01-0145-FEDER-032113. E.D.M. acknowledges the support by the Investigador FCT contract IF/00849/2015/CP1273/CT0003 and in the form of an exploratory project with the same reference. V.Zh.A. and S.G.S. also acknowledge the support from FCT through Investigador FCT contracts IF/00650/2015/CP1273/CT0001 and CEECIND/00826/2018 funded by FCT (Portugal) and POPH/FSE (EC). MT acknowledges the following grants: MIUR Premiale "Gaia-ESO survey" (PI S. Randich), MIUR Premiale "MiTiC: Mining the Cosmos" (PI B. Garilli), theASI-INAF contract 2014-049-R.O: "Realizzazione attività tecniche/scientifiche presso ASDC" (PI A. Antonelli), Fondazione Cassa di Risparmio di Firenze, progetto: "Know the star, know the planet" (PI E. Pancino). J.I.G.H. acknowledges financial support from the Spanish Ministry of Science and Innovation (MICINN) under the 2013 Ram\'on y Cajal program MICINN RYC–2013–14875 and from the Spanish MICINN project AYA2017-86389-P. We thank Chiaki Kobayashi for kindly providing the CGE models for carbon. Finally, we thank the anonymous referee, whose detailed review helped to improve this paper.\\

This research has made use of the SIMBAD database operated at CDS, Strasbourg (France) and the IRAF facility.

\end{acknowledgements}

\bibliographystyle{aa}
\bibliography{edm_bibliography}

\clearpage

\appendix

\section{Additional figures}

In this appendix we show the figures of [C/O] vs [O/H] using the forbidden oxygen line and the distribution of the elemental C/O ratios.\\

\vspace{2cm}

\begin{figure}
\centering
\includegraphics[width=8.7cm]{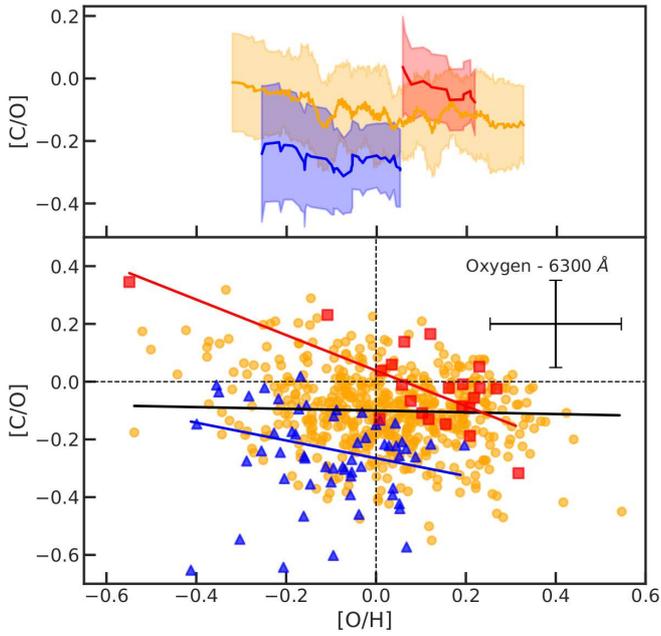}
\caption{[C/O] ratios as a function of [O/H] using the oxygen line at 6300\AA{} for the thin disk, thick disk and h$\alpha$mr stars. The running means were calculated using windows of 30, 15, and 10 for thin disk (yellow line), thick disk (blue line) and the h$\alpha$mr (red line), respectively. These numbers reflect the sizes of each sample. The 1$\sigma$ uncertainty in the moving mean is also shown as a shaded region of the corresponding colour.} 
\label{CO_OH_galaxy_pop_6300}
\end{figure}

\begin{figure}
\centering
\includegraphics[width=9.0cm]{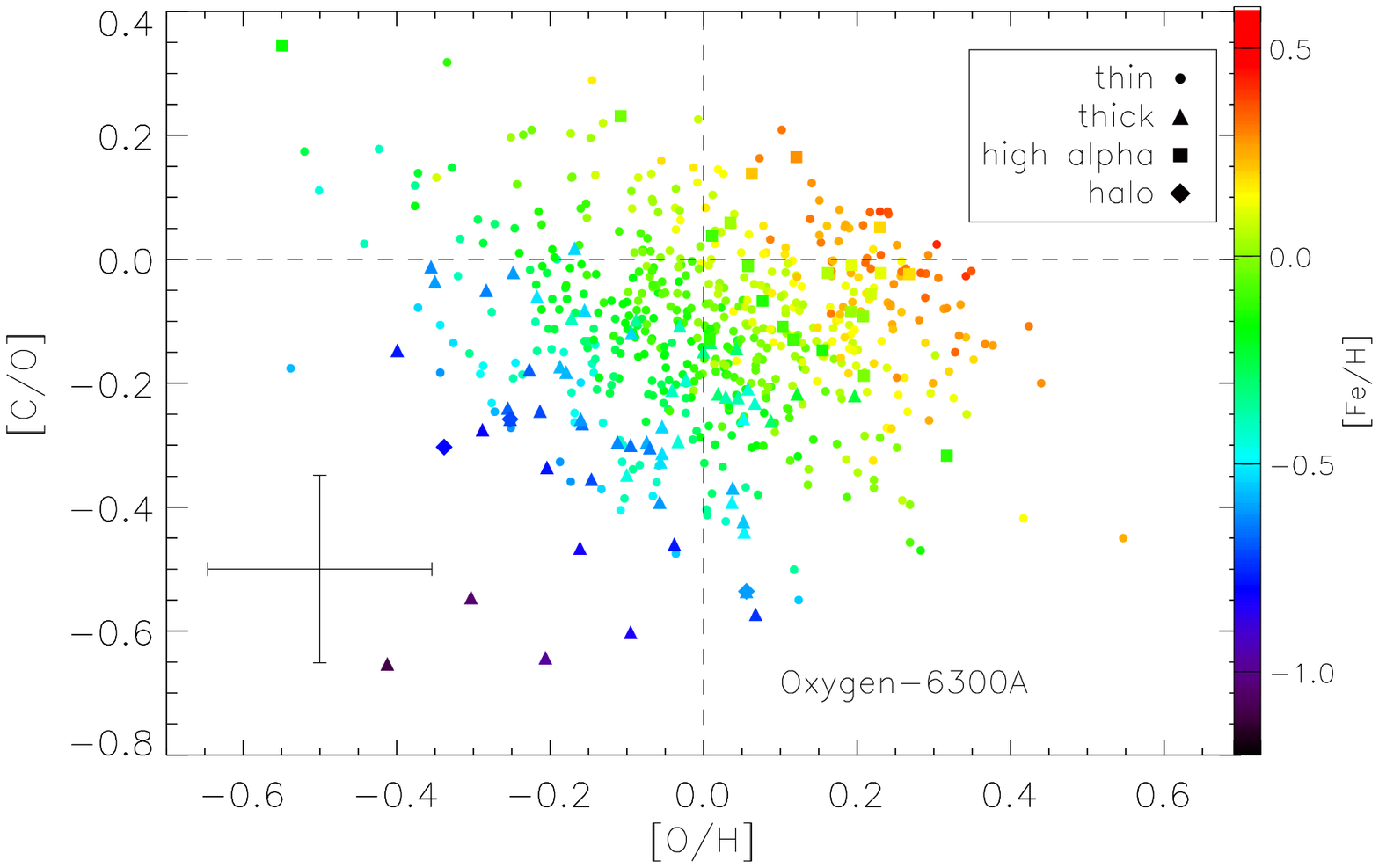}
\caption{[C/O] ratios as a function of [O/H] using the oxygen line at 6300\AA{} with a colour scale for metallicity.} 
\label{CO_OH_galaxy_6300}
\end{figure}

\begin{figure}
\centering
\includegraphics[width=9.0cm]{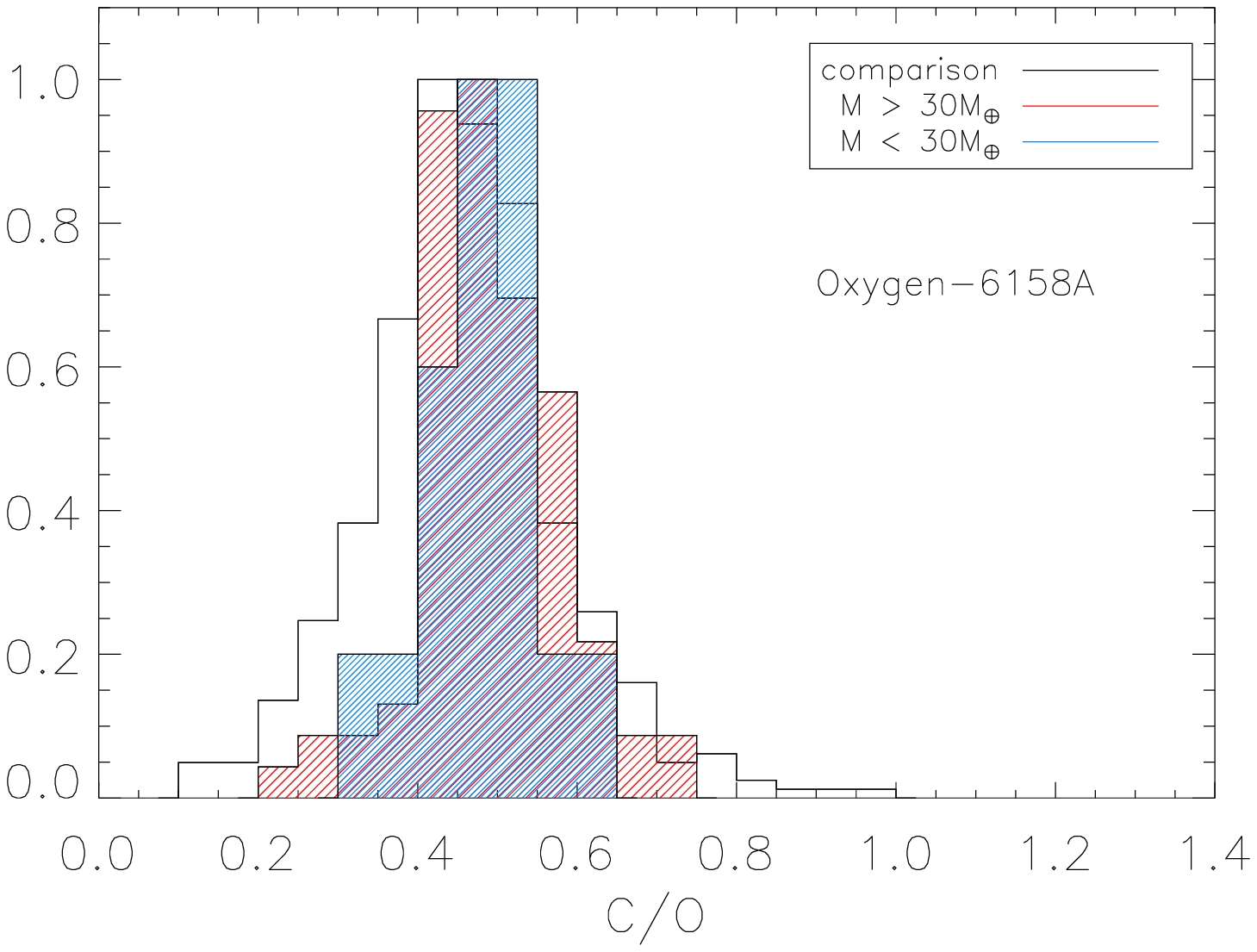}
\includegraphics[width=9.0cm]{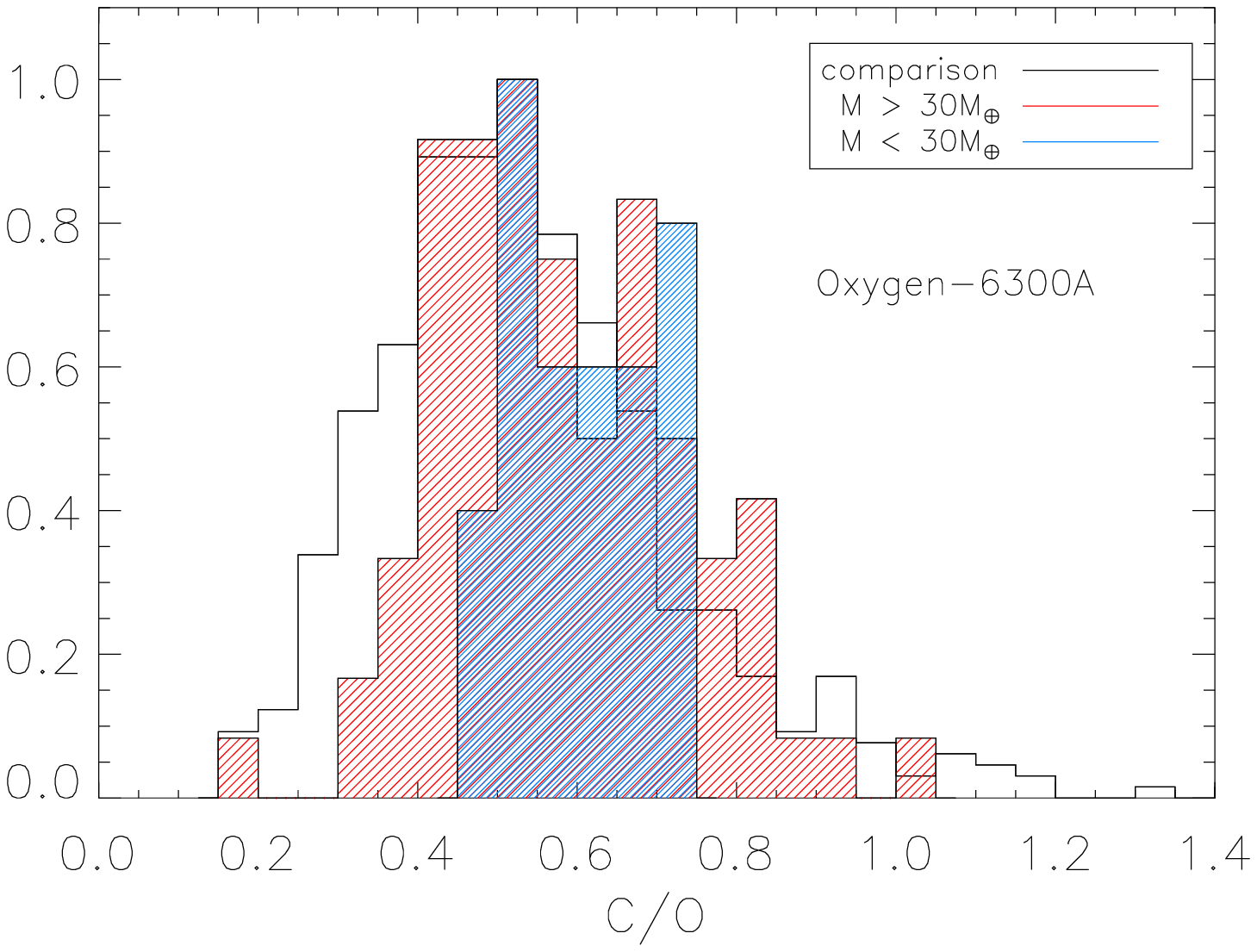}
\caption{Normalised distribution of C/O ratios for the populations of single stars, Neptunian hosts and Jupiter-type hosts, using the oxygen lines at 6158\AA{} (upper panel) or 6300\AA{} (lower panel).} 
\label{CO_histogram}
\end{figure}

\end{document}